\documentclass[useAMS,usenatbib]{mn2e}

\usepackage{amsmath,amssymb,mathrsfs,graphicx,float,latexsym}
\usepackage{epstopdf,ragged2e}
\usepackage{natbib}

\newcommand{\RA}{R_{\text{A}}}

\newcommand{\Ms}{M_{\star}}
\newcommand{\Rs}{R_{\star}}
\newcommand{\mus}{\mu_{\star}}
\newcommand{\Bs}{B_{\star}}
\newcommand{\Oms}{\Omega_{\star}}
\newcommand{\OmK}{\Omega_\text{K}}
\newcommand{\nuS}{\nu_\star}
\newcommand{\Is}{I_\star}
\newcommand{\oA}{\omega_{\text{A}}}
\newcommand{\NA}{N_{\text{A}}}
\newcommand{\Nm}{N_{\text{m}}}
\newcommand{\Rm}{R_\text{m}}
\newcommand{\Rco}{R_\text{co}}
\newcommand{\xA}{x_\text{A}}
\newcommand{\Ndisc}{N_\text{disc}}
\newcommand{\Ntot}{N_\text{tot}}
\newcommand{\nuQ}{\dot{\nu}_{\text{Q}}}
\newcommand{\nuO}{\dot{\nu}_{\text{O}}}

\newcommand{\be}{\begin{equation}}
\newcommand{\ee}{\end{equation}}
\newcommand{\bear}{\begin{eqnarray}}
\newcommand{\eear}{\end{eqnarray}}
\newcommand{\p}{\prime}

\newcommand{\nn}{\nonumber}

\newcommand{\bnabla}{\boldsymbol{\nabla}}

\newcommand{\cT}{{\cal T}}
\newcommand{\cC}{{\cal C}}

\newcommand{\rK}{{\rm K}}
\newcommand{\rrm}{{\rm m}}

\newcommand{\rA}{{\rm A}}

\def\apj{{ApJ}}

\def\apjl{{ApJL}}

\def\aap{{A\&A}}

\def\mnras{{MNRAS}}

\def\nat{{Nature}}

\def\prd{{Physical Review D}}
\def\prl{{Phys. Rev. Lett.}}

\def\04a{{2004 a}}
\def\04b{{2004 b}}

\begin{document}

\title[Modelling spin-up episodes in AMXPs]
{Modelling spin-up episodes in accreting millisecond X-ray pulsars}
\author[K.~Glampedakis \& A.G.~Suvorov]{Kostas Glampedakis$^{1,2}$\thanks{kostas@um.es} and Arthur G. Suvorov$^{2,3}$\thanks{arthur.suvorov@tat.uni-tuebingen.de}\\
$^1$Departamento de Física, Universidad de Murcia, Murcia, E-30100, Spain \\
$^2$Theoretical Astrophysics, Eberhard Karls University of T{\"u}bingen, T{\"u}bingen, D-72076, Germany \\
$^3$Manly Astrophysics, 15/41-42 East Esplanade, Manly, NSW 2095, Australia}

\date{Accepted ?. Received ?; in original form ?}

\pagerange{\pageref{firstpage}--\pageref{lastpage}} \pubyear{?}

\maketitle
\label{firstpage}

\begin{abstract}

\noindent Accreting millisecond X-ray pulsars are known to provide a wealth of physical information during their successive
states of outburst and quiescence. Based on the observed spin-up and spin-down rates of these objects 
it is possible, among other things, to infer the stellar magnetic field strength and test models of accretion disc flow. 
In this paper we consider the three accreting X-ray pulsars (XTE J1751--305, IGR J00291+5934 \& SAX J1808.4--3658) 
with the best available timing data, and model their observed spin-up rates with the help of a collection of standard torque models 
that describe a magnetically-threaded accretion disc truncated at the magnetospheric radius.
Whilst none of these models are able to explain the observational data, we find that the inclusion of the physically motivated 
phenomenological parameter $\xi$, which controls the uncertainty in the location of the magnetospheric
radius, leads to an enhanced disc-integrated accretion torque. These `new' torque models are compatible with the observed 
spin-up rates as well as the inferred magnetic fields of these objects provided that $\xi \approx 0.1-0.5$. Our results are supplemented 
with a discussion of the relevance of additional physics effects that include the presence of a multipolar magnetic field and 
general-relativistic gravity. 

\end{abstract}

\begin{keywords}
stars: neutron, magnetic fields, X-rays: binaries, accretion
\end{keywords}


\section{Introduction} 
\label{sec:intro}

Accreting neutron stars in low mass X-ray binaries (LMXBs) rank amongst the most well-studied compact objects in astrophysics. 
Interactions between the neutron star's strong magnetic field and general-relativistic (GR) gravity with the freely falling plasma accretion 
flow from the companion star can spin-up the neutron star to $\gtrsim$ millisecond rotation periods and provide fuel that can ignite atop 
the stellar surface, triggering thermonuclear explosions \citep{gk21}. As such, the rich physical environment of LMXBs can be used to study 
neutron star phenomenology in a variety of ways. For instance, the properties of X-ray flashes from bursting LMXBs \citep{li99,guv13}, and 
of thermal relaxation \citep{page13,pot18} or emissions \citep{bog19,nicer21} from quiescent systems, lead to measurements of local 
(e.g., crust microphysics) and global (e.g., mass-radius) quantities that are subsequently converted into constraints for the equation of state 
of neutron star matter \citep{latp01}. Mapping out the spin-temperature plane of LMXBs additionally allows one to study the intricate excitation 
and quenching mechanisms \citep{ho11,stroh14} of the gravitational wave-driven $r$-mode instability \citep{aks99}. 
Gravitational wave (GW) emissions from LMXBs could additionally take place as a result of the transient formation of quadrupolar `mountains'
and may be responsible for the observed spin-down irregularities \citep{haskell17}. The physics of LMXB accretion discs and magnetic fields
 -- the focus of this paper -- may also be revealed via quasi-periodic oscillations \citep{van06}, emission lines \citep{cack09}, and spin-up observations.

A particularly important subpopulation of LMXBs comprises the accreting millisecond X-ray pulsars (AMXPs) [see~\cite{amxp_review} for a review]; 
these systems emit X-ray pulses energised by the plasma captured from the accretion disc and channeled onto the neutron star's magnetic poles. 
The emission is modulated by the neutron star's rotation, thus allowing a precision measurement of the spin frequency. The hypothesised link between 
accreting neutron stars and the older population of recycled millisecond pulsars \citep{alp82,bhatt91} was confirmed by the discovery of the first AMXP 
in 1998 \citep{wij98} and the `swinging' pulsars IGR J18245--2452 and PSR J1023+2038, which alternate between radio- and X-ray loud states, some 
years later \citep{arch09,papp13}. Indeed, AMXPs could hardly be classified as steady-state systems; their accretion lifetime is punctuated by active 
phases during which a markedly increased mass accretion rate (which stands as a proxy for the observed X-ray luminosity) leads to a spin-up episode 
via the action of the accretion disc's torque. In between these episodes, the system accretes at a much lower rate (which may involve a tenuous receding 
inner disc) and the spin evolution in this state of quiescence is expected to be dominated by the star's own electromagnetic spin-down torque \citep{gunn69}. 

Much work has been dedicated to the study of the long-term, time-averaged spin equilibrium frequency of AMXPs (and of other 
LMXBs with known spin periods). Early models invoked a GW-accretion torque balance~\citep{bild98,levin99,andersson00} but subsequent 
work has shown that magnetic coupling to the accretion disc may be the key mechanism for spin equilibrium~\citep{rap04,andersson05,bc17}. 
More relevant to the present work is the theoretical modelling of spin-up episodes in AMXPs as observed during active outburst periods. 
Two of these systems, XTE J1751--305 and IGR J00291+5934, were studied by \cite{ajh14} with the help of standard 
accretion torque models available in the literature. They found that these torques fall short of explaining the spin-up 
and magnetic field data, thus casting some doubt on  our understanding of the physics of these systems. 

In this paper we revisit the topic of accretion spin-up in AMXPs by performing a systematic study of those three systems, 
XTE J1751--305, IGR J00291+5934 \& SAX J1808.4--3658, which currently have the best-measured spin evolution during outburst 
and quiescent phases. We provide a detailed discussion of the data and the associated observational and systematic uncertainties, 
and highlight some physical differences between the various sources. 
In the first part of the paper the spin-up torques are modelled within the framework of standard accretion theory, where the disc 
is coupled to the stellar magnetic field and is truncated at the magnetospheric radius. The new aspect of our approach, not considered by 
earlier work, lies in the incorporation of the phenomenological parameter $\xi$ as a measure of the uncertain physics in the vicinity of the 
magnetospheric radius. For the physically motivated case where $\xi < 1$, we find that models which account for magnetic field threading of the disc 
can lead to spin-up rates comparable to the observational data whilst being consistent with the inferred magnetic field from the spin-down data. 
This is the main result of this paper. 
The second part of our analysis consists of a quantitative discussion of what we consider to be the most important  `additional physics' corrections 
to the basic model, namely, a magnetic field which is not purely dipolar and some key effects of GR gravity. 

The rest of the paper is organised as follows. In Sections~\ref{sec:standardmodel}-\ref{sec:models} we discuss the various accretion 
torque models available in the literature and construct two `new' ones, which are then compared in Section~\ref{sec:Ncompare}. 
In Section~\ref{sec:Bcons} we discuss the constraints imposed on the magnetic field by the geometry of the accretion disc. 
The necessary spin evolution formulae are presented in Section~\ref{sec:spinevol}.
Section~\ref{sec:cases} is the main part of this paper and contains the comparison of the theoretical accretion torques
against the observed spin-up episodes of three AMXPs with reliable timing data (Sections~\ref{sec:XTE}-\ref{sec:SAX}). 
In Section~\ref{sec:morephysics} we resume our theoretical discussion of accretion physics by considering the effect
of a multipolar stellar magnetic field (Section~\ref{sec:multiB}) and of GR gravity (Sections~\ref{sec:GRcorrect}-\ref{sec:LT}). 
Our concluding remarks can be found in Section~\ref{sec:conclusions}. The Appendix contains some secondary technical details 
related to the structure of the standard accretion disc model. 

Notation:
Throughout the paper we use a star symbol to label stellar parameters. Moreover, we adopt the following fairly standard 
normalisations for the stellar mass $\Ms$, radius $\Rs$, and (dipole) polar field strength $\Bs$: $M_{1.4} = \Ms/1.4\,M_\odot$, 
$R_6 = \Rs/10^6\,\mbox{cm}$, and $B_8 = \Bs/10^8\,\mbox{G}$. In addition, the stellar spin frequency $\nuS$, accretion rate 
$\dot{M}$, and X-ray luminosities $L_{\rm X}$ are normalised as $\nu_{500} = \nuS/500\,\mbox{Hz}$,  
$\dot{M}_{-10} = \dot{M}/10^{-10} M_\odot \mbox{yr}^{-1}$, and $L_{\rm X,36} = L_{\rm X}/10^{36} \text{ erg s}^{-1}$, respectively.

\section{Models of accretion torques}
\label{sec:Models}

This theoretical first part of the paper provides a detailed survey of the various analytical accretion torque models available 
in the market. As discussed below, these are largely phenomenological constructions that correspond to different viable
choices for the disc's truncation radius and its interaction with the stellar magnetic field.  Using the same logic, we add 
one more model to this torque collection and then go on to compare these models against observed spin-up episodes of
AMXPs.

\subsection{Baseline accretion torque model with a magnetic field}
\label{sec:standardmodel}

In what could be called the `standard accretion torque model'  the disc is assumed to be geometrically thin
and quasi-stationary over timescales much longer than the hydrodynamical and orbital timescales of the 
accreted matter. The interaction of the stellar magnetic field with the disc is assumed to fall well within the
domain of the usual magnetohydrodynamic (MHD) framework. In addition, this baseline model assumes Newtonian gravity (the impact of
GR gravity is discussed in a later section) and axisymmetry with respect to the stellar spin axis.

The disc's `equation of motion' (in standard cylindrical coordinates $\{r,\varphi,z\}$) is the following thickness-integrated Euler equation
[cf. \cite{ghosh79b}, \cite{rap04}],
\be
-\dot{M} \frac{d}{dr}\left  [ \Omega (r) r^2 \right ] = B_z B_\varphi r^2 + \cT_{\rm visc},
\label{EulerMHD}
\ee 
where $\Omega(r)$ is the disc's rotational profile and $\cT_{\rm visc}$ is the viscous torque. 
The magnetic torque exerted on the disc by the stellar field comprises the poloidal and toroidal field components, $B_z$ and 
$B_\varphi$ respectively. The magnetic field is assumed dipolar (this assumption is relaxed in a later section where we include 
higher magnetic multipole moments),
\be
B_z = - \Bs \left ( \frac{\Rs}{r} \right )^3 = -\frac{\mus}{r^3},
\label{polfield}
\ee
where $\Bs$ is the surface polar field and $\mus$ is the corresponding dipole moment. 

The magnetic field is likely to play an important dynamical role during accretion and dominate the flow below a `magnetospheric' (or Alfv\'en) radius 
$\Rm$.  The ensuing physical picture is that of a disc truncated in the region $r \approx R_\rrm$  with accreted matter being entirely channeled along 
the field lines and onto the polar caps (note, however, that the situation may be far more complex for a `weak'  magnetic field). Across the same region 
the angular frequency $\Omega$ is assumed to make a smooth transition from a Keplerian profile, $\OmK (r)= \sqrt{G \Ms/r^3}$, to the stellar angular 
frequency, $\Oms$~\citep{rap04}.

The magnetospheric radius is the first key lengthscale of the present accretion model. It is common practice in the literature to 
estimate this parameter based on the assumption of comparable energy densities for the orbiting gas and the (poloidal) 
magnetic field in the disc's truncation region~\citep{accbook}. If we define as $\RA$ the resulting solution for $\Rm$, we have
\be
\frac{1}{2} \rho \OmK^2 \RA^2 \approx \frac{B^2_z}{8\pi}.
\ee
When combined with standard thin disc structure equations (these are listed in Appendix~\ref{sec:thindisc}) this relation leads to,
\be
\RA \approx \xi \frac{\mus^{4/7}}{ \dot{M}^{2/7} (G\Ms)^{1/7}}.
\label{RA}
\ee
This calculation's phenomenological parameter $\xi$ is defined as,
\be
\xi =  (6 \pi \alpha)^{2/7} \left (\frac{H}{R_\rA} \right )^{6/7} 
\approx 0.2  \left (\frac{\alpha}{0.1} \right )^{2/7} \left ( \frac{H/R_\rA}{0.1}\right )^{6/7},
\label{xinum}
\ee
where in the second equation we have normalised the disc's thickness $H$ and viscosity parameter $\alpha$ to their
`canonical' values. To some extent this parameter is a measure of our ignorance of the complicated physics taking 
place in the vicinity of the disc's truncation radius and it is typically assumed to vary within a range $\xi \approx 0.1-1$ 
(see also below). For $\RA$ itself we obtain the numerical estimate,
\be
\RA 
\approx 35\,\xi\, \dot{M}_{-10}^{-2/7} M_{1.4}^{-1/7} R_{6}^{12/7} B_8^{4/7}\,~ \mbox{km}.
\label{Rmnum}
\ee
Alternatively, the magnetospheric radius can be obtained via a direct application of Eq.~\eqref{EulerMHD}, after
setting $ \cT_{\rm visc} \approx 0$ at the disc's truncation radius and approximating  
$d[ \Omega r^2 ]/dr \approx \OmK \RA^2/\Delta r_\rrm$ where $\Delta r_\rrm$ is the radial width of the truncation
\citep{psaltis99}. The outcome of this calculation resembles Eq.~\eqref{RA} with 
\be
\xi = \left (\lambda_B \frac{\Delta r_\rrm}{\RA} \right )^{2/7}, \qquad  \lambda_{\rm B} \equiv \left | \frac{B_\phi}{B_z}  \right |_{\RA}.
\ee
\cite{psaltis99} assume $\lambda_{\rm B} \sim 1$ and estimate $ \Delta r_\rrm /\RA \sim 0.01-1 $
 which translates to $\xi \approx 0.3 -1$.
 
The two preceding (approximate) calculations clearly show that the $\xi$ parameter lumps together uncertainties related
to the disc structure as well as the relative poloidal-toroidal magnetic field strength in the vicinity of the magnetospheric radius.  

The second key lengthscale of any accretion torque model is the so-called corotation radius $\Rco$, defined as the radial distance
where the orbital and stellar frequencies match, i.e. $\Oms = \OmK (\Rco)$. From this we easily find,
\be
\Rco \approx 27\, M_{1.4}^{1/3} \nu^{-2/3}_{500}\,~\mbox{km}.
\ee
Based on the above estimates we should expect $ \RA \sim \Rco$.

The truncation of the disc at $r \approx R_\rA $ is associated with a `material'  Alfv\'en torque of increased lever-arm length
(as compared to that of a non-magnetic system)~\citep{pringle72},
\be
\NA=  \dot{M} \RA^2 \OmK (\RA) = \dot{M} \sqrt{G\Ms \RA}.
\label{Nm1}
\ee
Field lines rotating faster than the local Keplerian speed produce a negative torque and may lead
to a propeller effect when $\RA > \Rco$. In this regime the accretion flow will be centrifugally inhibited and 
matter may be ejected from the system [though see also \cite{taam93}]. As accreting matter is flung away, the star would experience a spin-down torque. 
A simple way to account for this effect is by modifying the previous torque~\citep{andersson05,ajh14},
\be
\Nm = \dot{M} \RA^2 [ \, \OmK (\RA) - \Oms \, ]  = \NA \left (1- \oA \right ),
\label{Nm2}
\ee
where we have introduced the so-called (dimensionless) fastness parameters,
\be
\oA \equiv \xA^{3/2}, \qquad \xA\equiv  \frac{\RA}{\Rco}.
\ee
The phenomenological expression \eqref{Nm2} predicts spin equilibrium, $\Nm=0$,  to take place at $\xA =1$ in accordance 
with the intuitive picture described above. 
This equality translates to the following equilibrium spin frequency,
\be
\nu_{\rm eq} \approx 283\, \xi^{-3/2}  B_8^{-6/7} R_{6}^{-18/7} \dot{M}_{-10}^{3/7} M_{1.4}^{5/7} \,~\mbox{Hz},
\label{feq1}
\ee
which is in good agreement with the average spin frequency of the known LMXB population [see e.g., \cite{patruno17}].

\subsection{Accretion torque with a magnetically-threaded disc}
\label{sec:models}

We can raise the sophistication  level of the preceding  baseline model by taking into account the magnetic 
field-disc coupling and the ensuing wind-up of the field lines by the orbiting matter~\citep{ghosh79b,wang95}. 
The generated toroidal field is described by the induction equation,
\be \label{induction}
\partial_t B_\varphi = | \bnabla \times  (\mathbf{v} \times \mathbf{B} ) |_\varphi.
\ee
This equation can be analytically handled by approximating $\partial_t B_\varphi \approx B_\varphi/\tau_\varphi$
and $\mathbf{v}= [\OmK (r) - \Oms ]\hat{\boldsymbol{\varphi}} $. The physics behind the timescale $\tau_\varphi$ is somewhat
sketchy; following \cite{wang95} (which provides the most detailed analysis on the subject) we can parametrise the toroidal field as
\be
B_\varphi (r) =  \zeta B_z (r) f [\Oms/\OmK (r) ],
\ee
where $\zeta$ is yet another phenomenological constant parameter. 
The function $f$ depends on the mechanism responsible for limiting the growth of $B_\varphi$; 
for turbulent diffusion in the disc (`mechanism (2)') and magnetic reconnection outside the disc (`mechanism (3)')
\cite{wang95} gives:
\be
f_{\rm (2)} = \frac{\Oms}{\OmK}  -1, 
\quad
f_{\rm (3)} = 
\begin{cases} 
f_{\rm (2)}, ~~\qquad r<\Rco \\
\\
1- \Omega_\rK/\Oms,~ r > \Rco 
\end{cases}
\label{Btor}
\ee
According to both prescriptions the toroidal field is generated in the prograde (retrograde) direction for 
$R < \Rco$ ($ R > \Rco$).

As a result of the magnetic field lines threading the disc there is an additional accretion torque $\Ndisc$ exerted on the
neutron star. This is given by the integral\footnote{Strictly speaking, the integral's upper limit should be set at the
light cylinder radius, $R_{\rm lc} = c/\Oms \approx 96\, \nu_{500} \,\mbox{km}$, which marks the separatrix of the
last closed magnetic field line. However, the error introduced by taking the integral out to infinity is negligible given that 
$\Rm, \Rco \ll R_{\rm lc}$.},
\be
\Ndisc = - \int_{R_\rrm}^\infty dr r^2 B_\varphi B_z. 
\ee
For the two toroidal field choices~\eqref{Btor} we find
\begin{align}
\Ndisc^{\rm (2)} &=  \frac{\zeta \mus^2}{3 \Rm^3} \left ( 1 - 2 \omega \right ),
\label{Ndisc2}
\\
\nn \\
\Ndisc^{\rm (3)} &=  \frac{\zeta \mus^2}{9 R_\rrm^3} \left ( 3 -6 \omega + 2 \omega^2 \right ),
\label{Ndisc3}
\end{align}
where we have defined a new pair of fastness parameters,
\be
\omega \equiv x^{3/2},  \quad x \equiv \frac{\Rm}{\Rco},
\ee
The total accretion torque is the sum of the disc-integrated torque $\Ndisc$ and the material torque at $r=\Rm$. 
This is given by the earlier baseline expressions~\eqref{Nm1}, \eqref{Nm2} with $\RA$ replaced by a general $\Rm$
magnetospheric radius.

The reason we have allowed for the possibility of $\Rm \neq \RA$ in this section is that Eq.~\eqref{Btor}, in combination with the 
assumption $\Omega(r) =\OmK(r)$, allows the Euler equation~\eqref{EulerMHD}  to become a relation for $\Rm$,
\be
\Rm = (2\zeta)^{2/7} \frac{\mus^{4/7}}{ \dot{M}^{2/7} (G \Ms)^{1/7}} \left [\, 1 - \left (\frac{\Rm}{\Rco} \right )^{3/2} \,\right ]^{2/7}. 
\label{Rm2}
\ee
As is evident, this expression is self-consistent provided $\Rm < \Rco$. Moreover, it reduces to the earlier Alfv\'en radius~\eqref{RA},
for $\Rm \ll \Rco$ and $  \xi \to  (2\zeta)^{2/7}$. 

With this identification between phenomenological parameters, we can rewrite~\eqref{Rm2} as
\be
 x = x_\rA \left ( 1 - \omega \right )^{2/7},
 \label{xxA}
\ee
and we can see that apart from $x \leq 1$ we should also expect $x < x_\rA$ (i.e. $\Rm < \RA$). 

The above  torques and magnetospheric radii are combined in different ways in different papers in the literature.
For example, in \cite{wang95}  the total torque is given by, 
\be
\Ntot^{\rm W (2,3)} =   \dot{M} \sqrt{G \Ms \Rm}   + \Ndisc^{\rm (2,3)},
\label{Ntot2}
\ee
which with the further input of~\eqref{xxA} leads to,
\begin{align}
\Ntot^{\rm W(2)} &=   \frac{1}{3} \NA \frac{(7/2 -4\omega)}{(1- \omega)^{6/7}} ,
\\
\nn \\
\Ntot^{\rm W(3)} & =   \frac{1}{3} \NA  \frac{\left [\,7/2-4\omega + (1/3) \omega^2 \, \right ]}{(1-\omega)^{6/7}}. 
\end{align}
These total torques predict spin equilibrium at
\be
x_{\rm eq}^{(2)} \approx 0.91, \qquad x_{\rm eq}^{(3)} \approx 0.97.
\ee
 \cite{andersson05} adopt the magnetospheric radius of Eqs.~\eqref{Rm2}, \eqref{xxA} but opt for the material torque~\eqref{Nm2}
in combination with the mechanism (2) disc torque. The resulting total torque is,
\begin{align}
\Ntot^\rA &= \dot{M} \sqrt{G \Ms \Rm} (1-\omega)  + N_{\rm disc}^{\rm (2)}
\\
& =  \frac{1}{3} \NA  \frac{(\, 7/2 - 7 \omega + 3\omega^2 \, )}{(1- \omega)^{6/7}}.
\end{align}
The corresponding spin equilibrium is found to be,
\be
x_{\rm eq}^\rA \approx 0.81.
\ee
Finally, \cite{rap04} and \cite{bc17} use the mechanism (3) torque of \cite{wang95} with $\zeta=1$ but deviate from 
that model by adopting the magnetospheric radius $\Rm = \RA$ of the baseline model with $\xi=1$. 
The resulting torque is,
\be
\Ntot^{\rm R, BC} =  \NA + \Ndisc^{\rm (3)} 
=  \frac{2}{3}  \NA \left (2 - \oA + \frac{1}{3} \oA^2 \right ).
\label{NtotBC}
\ee
In contrast to the previous cases this torque does \emph{not} admit a point of equilibrium.

It is straightforward to invent two `new'  torque models based on the above mechanism (2) \& (3) 
prescription  and by choosing $\Rm = \RA$ \emph{without} assuming $\xi=1$.  The first of these
torques generalises expression~\eqref{NtotBC},
\be \label{eq:eqfast}
\Ntot^{\rm new (3)} =  \frac{2}{3}  \xi^{-7/2} \NA \left ( \frac{1+ 3\xi^{7/2}}{2} - \oA + \frac{1}{3} \oA^2 \right ).
\ee 
This torque vanishes at
\be
\omega_{\rm A, eq}^{\rm new(3)} = \frac{3}{2} \left (1 - \sqrt{\frac{1-6\xi^{7/2}}{3}} \right ),
\ee
which is real-valued for $\xi < \xi_{\rm max} \approx 0.6$. If we assume $0.1 < \xi < \xi_{\rm max}$, the corresponding
equilibrium $x$-point lies within the range,
\be
0.74 \lesssim x_{\rm A, eq}^{\rm new(3)} \lesssim1.3.
\label{xAeqn3}
\ee
The second new torque is
\be
\Ntot^{\rm new(2)} = \frac{1}{3} \xi^{-7/2} \NA \left (1 + 3 \xi^{7/2} - 2\oA \right),
\ee
and the associated equilibrium fastness parameter is
\be
\omega_{\rm A, eq}^{\rm new(2)} = \frac{1}{2} \left (  1+ 3 \xi^{7/2} \right ).
\ee
For  the $0.1 < \xi < 1$ range this result returns,
\be
0.63 < x_{\rm A, eq}^{\rm new(2)} \lesssim 1.6.
\label{xAeqn2}
\ee
We can notice that both new models can accommodate $\RA > \Rco$ (i.e. the propeller regime of the baseline model) 
as a viable spin-up regime.


\subsection{Comparing the various torques}
\label{sec:Ncompare}

None of the previous torques can be classified as `rigorous' but, nevertheless, they do represent the state-of-the-art  
when it comes to modelling the spin evolution of accreting neutron stars.  Among the models discussed we should 
expect those with a magnetically-threaded disc to be the most realistic ones. The situation is less clear 
when choosing\footnote{A rather different estimate for $\Rm$ comes from 3D MHD simulations of plasma flow \citep{kulk13}
in the form of a fitting formula $\Rm \approx  ( \Rs^{3} \mus^4 / \dot{M}^2 G \Ms)^{1/10}$ that takes into account the 
non-dipolar deformation of the neutron star's magnetosphere.}
between $\RA$, Eq.~\eqref{RA}, or $\Rm$, Eq.~\eqref{Rm2}. The first expression is the most widely
used in the literature, although typically served with $\xi=1$  which, in face of the estimate~\eqref{xinum}, may not
be fully justified. Meanwhile, the assumption of a Keplerian angular frequency in the derivation of Eq.~\eqref{Rm2} 
may be equally unrealistic. 

To some extent this discussion boils down to choosing the unknown function in the Euler 
equation~\eqref{EulerMHD}. As we have seen, this equation can be approximated with respect to 
$\Omega$ and $B_\varphi$ and solved for $\Rm=\RA$~\citep{psaltis99}; it can be solved for $\Rm$ assuming
$\Omega=\OmK$ and a specific functional form for $B_\varphi(r)$ \citep{wang95}; it can be solved for $\Omega(r)$ 
for a given  functional form $B_\varphi (r)$ and $\Rm=\RA$ or for the viscous stress $\cT_{\rm visc}(r)$ 
after assuming a Keplerian angular frequency \citep{kluz07}.
 
\begin{figure*}
\hspace{2cm} \includegraphics[width=0.6\textwidth]{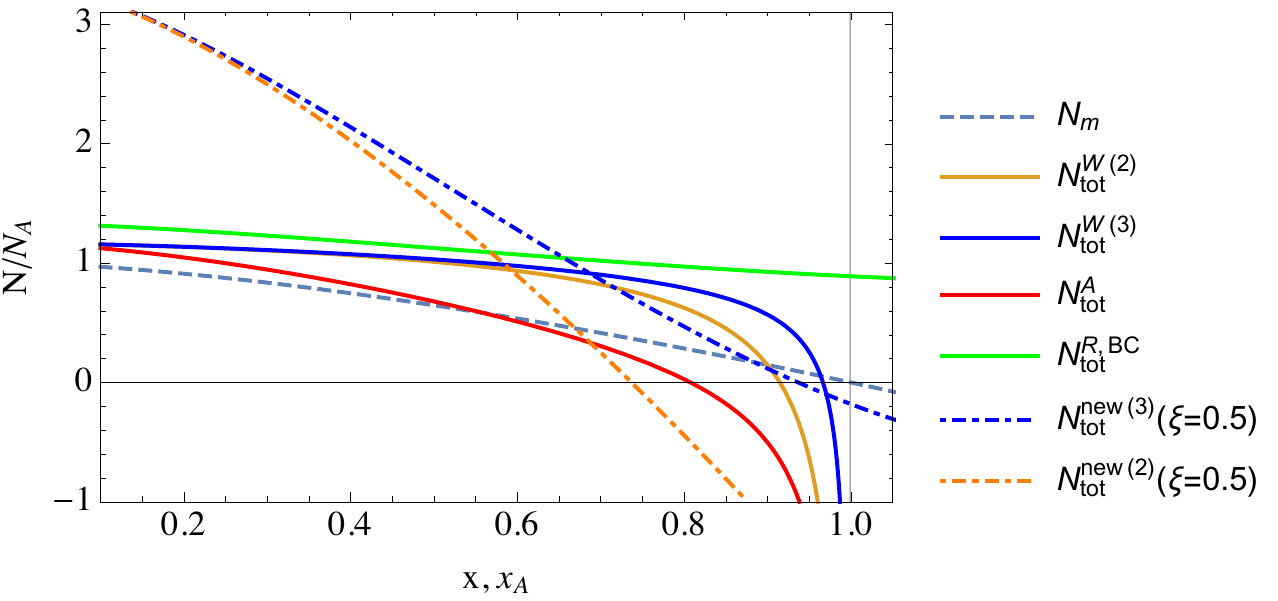}
\caption{Comparison of the torques pigeonholed in Table~\ref{tab:accmodels} and discussed in Section~\ref{sec:models}. 
Each torque is normalised to the baseline torque $\NA$ and is plotted as a function of the corresponding fastness parameter $\xA$ or $x$. 
When plotting the two `new' torques $\Ntot^{\rm new(2,3)}$ we factor out the $\xi$ dependence of $\NA$ and then evaluate the torque at the 
fiducial value $\xi=0.5$. $N=0$ marks the equilibrium $x$-point of each model (only the baseline torque $\Nm$ predicts $x_{\rm eq}=1$, see vertical line).}
\label{fig:torques}
\end{figure*}

As an executive summary, the various accretion torques that have been suggested in the literature (plus the ones discussed here) 
are listed in Table~\ref{tab:accmodels}. Their relative strength (normalised to $\NA$) as a function of the fastness parameter 
$x$ (or $\xA$) is shown in Fig.~\ref{fig:torques}.  We can see that among the previously used torque models, $\Ntot^{\rm R, BC}$ is the 
dominant one across the entire fastness parameter range. But the most prominent feature in this plot is the enhanced strength 
of the pair of `new' torques as a result of the $\xi <1$ degree of freedom. In particular, the enhancement originates from the 
negative $\xi$-power dependence of the disc-integrated portion $\Ndisc$ of the torque rather than the material part $\NA$.
\begin{table*}
\begin{tabular}{ccc}
 \hline
 \hline
Torque symbol & Functional form & Reference(s) \\
\hline
$\NA$                         & $ \dot{M} \sqrt{ G \Ms \RA}$ & \citep{pringle72} \\
 && \\
$ \Nm$                        & $\NA \left( 1 - \oA \right)$ & [cf. \cite{andersson05}] \\
 && \\
$\Ntot^{\text{W(2)}} $  & $ \frac {1} {3} \NA \left( 7/2 - 4 \omega \right)  \left( 1 - \omega \right)^{-6/7}$ & \citep{wang95} \\
 && \\
$\Ntot^{\text{W(3)}} $  & $\frac {1} {3} \NA  \left(7/2 - 4 \omega + \omega^2/3\right) \left( 1 - \omega \right)^{-6/7}$ & \citep{wang95} \\
 && \\
$\Ntot^{\text{R,BC}}$  & $\frac {2} {3} \NA \left(2 - \oA + \frac {1} {3} \oA^2 \right)$ & \citep{rap04} \\
 &   & \citep{bc17} \\     
 && \\             
$\Ntot^{\text{A}} $       &  $\frac {1} {3} \NA  \left(7/2 - 7 \omega + 3 \omega^2 \right) \left( 1 - \omega\right)^{-6/7}$ & \citep{andersson05} \\
 && \\
$\Ntot^{\text{new(3)}} $  & $\frac {2} {3} \xi^{-7/2} \NA \left[ \, \frac{1}{2} \left ( 1 + 3 \xi^{7/2} \right ) - \oA + \frac {1} {3} \oA^2 \,\right]$ & This paper \\
 && \\
 $\Ntot^{\text{new(2)}} $  & $  \frac{1}{3} \xi^{-7/2} \NA \left (1 + 3 \xi^{7/2} - 2\oA \right) $ & This paper \\
\hline
\hline
\end{tabular}
\caption{A list of the accretion torques considered in this paper (see Section~\ref{sec:models}). }
\label{tab:accmodels}
\end{table*}
%
%

It is worth pointing out that our discussion of the various torque models presented in Table~\ref{tab:accmodels}  may
be of some interest to the still open question of the spin distribution of AMXPs. The suggested separation in `fast' and 
`slow' subpopulations~\citep{patruno17} could reflect the operation of different torque mechanisms and/or different
$\xi$-parameter physics that drive different systems to different spin equilibria as in Eqs.~\eqref{xAeqn3} and \eqref{xAeqn2}.


\subsection{Constraints on the stellar magnetic field}
\label{sec:Bcons}

The basic assumptions underpinning the standard accretion model discussed in the
preceding sections places some constraints on the stellar magnetic field.

The first constraint comes from the condition $\RA >\Rs$ (i.e. the truncation of the disc takes place
above the stellar surface) and leads to a lower limit for $\Bs$
\be
\Bs >  B_{\rm min} =  \frac{ \dot{M}^{1/2}(G\Ms)^{1/4}}{\xi^{7/4} \Rs^{5/4}}.
\label{Blow}
\ee
In a similar fashion, an upper limit for the magnetic field can be derived from the condition $ \xA <  x_{\rm A, eq}$ 
(i.e. $\Ntot >0$ during a spin-up episode),
\be
 \Bs  <  B_{\rm max} =\frac{ x_{\rm A, eq}^{7/4}}{(2\pi)^{7/6} \xi^{7/4}}  
 \frac{\dot{M}^{1/2} (G\Ms)^{5/6}}{ \nuS^{7/6} \Rs^3}.
 \label{Bhigh}
\ee
For a given $\Bs$ the above inequalities can be rearranged into constraints for $\xi$; these should be considered 
together with this parameter's previously discussed theoretical range (for example, in the $N_{\rm tot}^{\rm new (3)}$ torque model $\xi$ should like below 
$\xi_{\rm max}$ for the system to be able to reach spin equilibrium). 
This issue is discussed further in a later section for specific cases of AMXPs.

Similar constraints can be derived for the magnetospheric radius $\Rm$ described by~\eqref{Rm2}, 
although via a numerical solution.

\subsection{Spin evolution}
\label{sec:spinevol}

Given a total torque $N$, the (instantaneous) spin evolution of the system, $\dot{\nu}_\star$, is determined by balancing 
the shift in the rotational kinetic energy with that associated with $N$, i.e.,
\be
\label{eq:torquebal}
\dot{\nu}_\star = \frac {N} {2 \pi \Is},
\ee
where we have assumed that the moment of inertia is time independent, $\dot{I}_\star = 0$. Even during an active phase, 
the neutron star within the system will be subject to an electromagnetic braking torque, $N_{\text{EM}}$, that acts to slow down 
the star. For a given accretion torque $N_{\text{acc}}$ (i.e., one of those appearing in Tab. \ref{tab:accmodels}), the total torque 
acting on the star is thus $N = N_{\text{acc}} - N_{\text{EM}}$ (though see below). 

During active phases where the source is especially bright, one expects that $N_{\text{acc}} \gg N_{\text{EM}}$, and so the 
braking term is ignored when modelling the spin evolution during an outburst. During a period of quiescence however, where 
negligible accretion torques are applied, the star will decelerate. We adopt the subscript `Q'  throughout to indicate 
quiescent phase predictions or measurements. By contrast, `O' is similarly used to indicate an outburst phase.

Assuming pure (centred) dipole magnetic braking, the associated spin-down rate is given by a formula of the form
\be
\dot{\nu}_\text{Q} = - \frac{2\pi^2}{c^3} \frac{\mus^2 \nuS^3}{\Is}  K.
\label{eq:dipnu}
\ee
The parameter $K$ encompasses different choices of spin-down model (we also note the factor $1/2$ difference 
between our definition $\mus = \Bs \Rs^3$ and that of the two papers cited below). 
The classic vacuum model~\citep{gunn69} corresponds to $K = (1/3) \sin^2 \vartheta$, where $\vartheta$ is the spin-magnetic
axis misalignment angle. However, the standard practice of assuming a star that acts as an orthogonal rotator
($\vartheta=\pi/2$) does not sit well with the assumed axisymmetry of the accreting system. 
A more realistic approach would be to use the \cite{spit06} formulae [see also \cite{phil15}], appropriate for an oblique rotator coupled 
to a force-free magnetosphere, as in \cite{ajh14}. This state-of-the-art model is described by $K\approx (1/2) ( 1+ \sin^2\vartheta)$ 
and has the attractive property of predicting spin-down even for an aligned rotator.

We further note that expression \eqref{eq:dipnu} assumes negligible gravitational radiation and propeller torques. As such, the magnetic field 
estimates we obtain from spin-down should be treated as strict \emph{upper limits} 
(the same estimates would instead represent lower limits if the quiescence phase were to include an unmodelled spin-up torque 
e.g. due to weak residual accretion). 
For example, should the star house a time-dependent mass 
quadrupole moment through the formation of an accretion-built mountain, radiation reaction will sap additional angular momentum from the system 
and a lower $B_{\star}$ would be required to accommodate a fixed $\dot{\nu}_{\text{Q}}$ \citep{mp05,pri11}. Numerical simulations of Ohmic \citep{vig09} 
and thermal \citep{suvm19} relaxation suggest sufficiently light mountains can survive over long diffusion timescales ($\tau_{\text{diff}} \gtrsim 10^{5}$ yr), 
and may therefore persist during quiescent phases if formed during a previous, active epoch [though cf. \cite{mukh17}]. Because the GW 
torque scales sharply with the spin frequency, $N_{\text{GW}} \propto \nu^{5}$, it has been suggested that bimodality in the distribution of spins in AMXPs, 
distinguishing the `fast' and `slow' populations, could be explained by the presence or absence of GW torques \citep{ga19}. We will however 
assume that gravitational radiation is negligible for the remainder of this work, as these issues are beyond the scope of this paper [see, e.g., \cite{patruno10} 
for a discussion].

\section{Case studies: XTE J1751--305, IGR J00291+5934 \& SAX J1808.4--3658}
\label{sec:cases}

Using the torque models derived in the previous section, we are now in a position to compare theory with observation. 
Though we compile data relevant for several systems for completeness (see below), we focus primarily on three AMXPs in this paper: 
XTE J1751--305 (henceforth J1751), IGR J00291+5934 (J00291), and SAX J1808.4--3658 (J1808). Each of these sources have exhibited 
at least one well-timed outburst, where pulse timing revealed an increase in the respective spin frequencies between before and after phases. 
The first two of these were considered by \cite{ajh14}, who compared the baseline $\NA$ and $\Nm$ torques against the observations, and by 
considering these systems in detail we can provide something of a one-to-one comparison with their results. In particular, as commented by~\cite{ajh14}, 
existing models of spin-up were unable to explain the measured $\dot{\nu}_{\text{O}}$ values for these two objects during their 
respective 2002 and 2004 outbursts (see Fig. 2 therein). As will become clear, the discrepancy is even more extreme for J1808, which underwent 
a particularly violent episode of spin-up in 2015 \citep{sanna17}. One goal of this work is to review the findings of \cite{ajh14} using the various 
models presented in Tab. \ref{tab:accmodels}.

Table \ref{tab:nsdata} lists data relevant for several AMXP systems. In particular, the second column gives the spin frequencies and the third lists 
the (mean) \emph{spin-downs} measured during quiescence (though sometimes long-term averages are taken without excluding outburst rises), 
from which we estimate the polar field strength using expression \eqref{eq:dipnu} (third column). The final two columns list the mean \emph{spin-up} 
and X-ray luminosities recorded during the relevant bursting episode, respectively. Some notes relevant to data for the individual systems listed in 
Tab. \ref{tab:nsdata} and the general observational methodologies are given throughout the remainder of this section.

\subsection{Observational and systematic uncertainties}
\label{sec:obs}

The methods used to estimate changes in spin frequency, either during a bursting episode or in some quiescent epoch, vary in the literature. 
In the simplest timing method, one notes that the spin frequency changes by an amount $\Delta \nu$ during a time window of length $\tau$, which 
implies a mean spin-up of $\Delta \nu / \tau$. Often, however, a more sophisticated timing model using Taylor expansions is employed; see, e.g., \cite{papp08}. 
Either way, typically only mean values for the spin-up and down can be reliably measured, and it is for this reason that averages are presented in 
Tab. \ref{tab:nsdata} above. From a modelling perspective, one must effectively consider a time-averaged version of equation \eqref{eq:torquebal}. 
Other complicated factors also play a role, such as the choice for the flux cutoffs where the burst is said to have concluded and handling correlations 
between the X-ray flux, pulse phases, and timing noise; compare, for instance, the spin-ups reported for the 2004 outburst of J00291 between 
\cite{falanga05} [$\langle \nuO \rangle = 8.4(6) \times 10^{-13} \text{ Hz s}^{-1}$] and \cite{patruno10} [$\langle \nuO \rangle = 5.1(3) \times 10^{-13} \text{ Hz s}^{-1}$]. 
Reported spin-downs are also average values obtained from timing over long observational windows, usually several years; compare the quiescent spin-downs 
reported for J00291 between \cite{patruno10} [$\langle \nuQ \rangle = -3.0(8) \times 10^{-15} \text{ Hz s}^{-1}$] and later by \cite{papp11} 
[$\langle \nuQ \rangle = -4.1(12) \times 10^{-15} \text{ Hz s}^{-1}$]. For concreteness, the latter results are employed in this work. In some cases 
(most notably J1808) different estimates for the spin-down immediately following outbursts are recorded; see Sec. \ref{sec:SAX}. 

From a calibration perspective, accurately determining the peak (and mean) X-ray luminosities for the objects listed here requires one to `correct' the raw flux data. Radiation is scattered and absorbed by the interstellar medium en route to the detector(s), resulting in the instrument reporting a lower flux than is truly being emitted, the extent of which depends on the (spatially-varying) hydrogen column density; see \cite{latt14} for a detailed discussion. Furthermore, X-ray burst emissions are generally composed of thermal components, originating from the stellar surface and possibly the disc, and a scattered Compton component, originating from some height above the surface \citep[e.g.,][]{keek18}. Models aiming to account for these effects differ slightly in the literature \citep[see, e.g., the Comptonization model of][]{gier05}, resulting in different (post-processed) light curves. Finally, determining the mean luminosity from a given light curve can itself be subject to model variability. In many cases, the flux observed from bursts tends to show an exponential decay, and therefore one can --- assuming that the X-ray flux is a good tracer of $\dot{M}$ --- write the mass accretion rate during a burst as $\dot{M}(t) \sim \dot{M}_{\text{peak}} \exp \left[ \left(t - T_{0} \right) / \tau \right]$ \citep{burd06}, where $\tau$ is the characteristic e-folding decay time and $T_{0}$ is some reference time. From this expression one finds the mean, $\langle \dot{M} \rangle \sim 0.63 \dot{M}_{\text{peak}}$.

Assumptions on the efficiency of the system can also play a role, as there is some energy lost in converting between the `accretion' and X-ray luminosities. 
Generally, a factor of $\sim$ 1--2 is accounted for in this respect, but different authors consider different factors.  Furthermore, assumptions on the distance 
and whether the source is radiating isotropically or narrowly beaming both affect the estimate for the true, bolometric luminosity [see equation (9) in \cite{ng21}]. 
Finally, it is important to note that magnetic field and torque estimates scale with the stellar mass and radius in various ways. For instance, in expression 
\eqref{eq:dipnu} one sees that $B_{\star} \propto \Ms^{1/2} \Rs^{-2}$ for fixed $\nuQ$. While it is traditional to take the canonical values $\Ms = 1.4M_{\odot}$ 
and $\Rs = 10$ km, these may not be appropriate for all LMXB systems\footnote{According to the latest results from the Neutron Star Interior Composition 
Explorer (NICER), a more realistic radius for a star with a canonical mass $\Ms = 1.4 M_{\odot}$ is $\Rs \approx 12$ km \citep{nicer21}. A $\sim 20\%$ 
increase in $\Rs$ can have a non-negligible effect in expressions that scale strongly with radius, such as \eqref{eq:dipnu}.}. 
For instance, using the 1998 outburst data for J1808, \cite{li99} found that the neutron star may be very compact; for $\Rs = 10$ km, 
the \emph{minimum} mass they estimate is $\Ms \approx 2 M_{\odot}$. Since $L_{\rm X} \sim G \Ms \dot{M} / \Rs$, taking instead a value $\Ms = 2 M_{\odot}$
leads to a $\lesssim 40\%$ \emph{decrease} in the inferred accretion rate $\dot{M}$, which is the relevant quantity appearing within the baseline torque $N_{\text{A}}$. 
We additionally assume throughout that  $\Is \approx 0.38 \Ms \Rs^2$, in accord with the GR calculation for a Tolman-VII equation of 
state with a star of canonical compactness \citep{latp01}. 

\begin{table*}
\caption{Observed and derived properties related to the five AMXPs considered in this work. The polar field strength, $\Bs$, is derived 
from the given (mean) quiescent spin-down rate $\langle \nuQ \rangle$ through the braking formula \eqref{eq:dipnu} with $K = ( 1 + \sin^2 \vartheta)/2$, 
where the given uncertainties incorporate the range $0 \leq \vartheta \leq \pi/2$. The magnetic field 
estimate is made assuming $\Ms = 1.4 M_{\odot}$ and $\Rs = 10 \text{ km}$, where we use the Tolman-VII moment of inertia, 
$\Is \approx 0.38 \Ms \Rs^2$. X-ray luminosities are computed from `corrected' fluxes (see text), where the assumed distance is given 
explicitly through the subscripts (e.g., $d_{8.5}$ indicates that a distance of $d = 8.5$ kpc was assumed to convert between fluxes and luminosities). 
See text for comments on individual source data.}
\begin{tabular}{cccccc}
 \hline
 \hline
Source & $\nu_{\text{spin}}$ (Hz) & $\langle \dot{\nu}_{Q} \rangle$ (Hz $\text{s}^{-1}$) & $B_{\star} (\times 10^{8} \text{ G})$& $\langle \dot{\nu}_{O} \rangle$ (Hz $\text{s}^{-1}$) & $\langle L_{\rm X} \rangle$ ($\times 10^{36}$ erg $\text{s}^{-1}$) \\
\hline
XTE J1751--305 (2002)$^{\text{a}}$ & 435.3 & $-5.5(12) \times 10^{-15}$ & $3.8(10)$ & $3.7(10) \times 10^{-13}$  & $\sim 17 \times d^2_{8.5}$ \\
\hline
IGR J00291+5934 (2004)$^{\text{b}}$ & 598.9 & $-4.1(12) \times 10^{-15}$ &  $2.0(7)  $ & $5.1(3) \times 10^{-13}$  & $\sim 2.7 \times d^2_{4.2}$ \\ 2015 Outburst$^{\text{c}}$ &  & &  & $3(5) \times 10^{-12}$ & $\lesssim 0.72 \times d^{2}_{4.2}$ \\
\hline
SAX J1808.4--3658 (1998)$^{\text{d}}$ & 401.0 & $-5.5(12) \times 10^{-16}$ &  $1.3(4) $ & $< 2.5 \times 10^{-14}$ & $\lesssim 5.1 \times d_{3.5}^2$ \\ 2002 Outburst$^{\text{e}}$ & & $-7.6(15) \times 10^{-14}$ &  $16(4) $ & $4.4(8) \times 10^{-13}$  & $\sim 6.3 \times d_{3.5}^2$ \\ 2015 Outburst$^{\text{f}}$ & & $-1.5(2) \times 10^{-15}$ & $2.2(5)$  & $2.6(3) \times 10^{-11}$ & $\lesssim 2.6 \times d_{3.5}^{2}$  \\
\hline
XTE J1814--338 (2003)$^{\text{g}}$ & 314.4 & $\sim -8.7(4.2) \times 10^{-15}$ & $\sim 8.2(32)$ & $<1.5 \times 10^{-14}$ & $\lesssim 2.2 \times d_{8}^2$ \\
\hline
IGR J17494--3030 (2020)$^{\text{h}}$ & 376.1 & $-2.1(7) \times 10^{-14}$ &  $9.2(30) $ & $<1.8 \times 10^{-12}$ & $\gtrsim 1.1 \times d_{10}^{2}$ \\
\hline
\hline
\end{tabular}
\footnotesize{\textbf{References:} $\vphantom{\text{a}}^{\text{a}}$ \cite{gier05,papp08,rigg11}. $\vphantom{\text{b}}^{\text{b}}$ \cite{falanga05,patruno10,papp11}. $\vphantom{\text{c}}^{\text{c}}$ \cite{tudor17,sanna17b} [though cf. \cite{defalc17}]. $\vphantom{\text{d}}^{\text{d}}$ \cite{hart09,hask11}. $\vphantom{\text{e}}^{\text{e}}$ \cite{burd06} [though cf. \cite{chak03}]. $\vphantom{\text{f}}^{\text{f}}$ \cite{sanna17,tudor17}. $\vphantom{\text{g}}^{\text{g}}$ \cite{krauss05,hask11,bag13}. $\vphantom{\text{h}}^{\text{h}}$ \cite{ng21}.}
\label{tab:nsdata}
\end{table*}


\subsection{XTE J1751--305}
\label{sec:XTE}

Table \ref{tab:nsdata} reports data relevant for the 2002 outburst of J1751. Before presenting a detailed comparison between torque models, we note that the 
peak luminosity given by \cite{rigg11} [$L_{\rm max, 36} \sim 11.6\, d_{8.5}^2$; the value used by \cite{ajh14}] is smaller than that of \cite{gier05} 
($L_{\rm max,36} \sim 27 \,d_{8.5}^2$) [see also \cite{papp08}]. These latter references include a Comptonization component, which suggests the true (bolometric) 
luminosities are roughly twice as large as the raw values, which is likely the cause of the discrepancy. The distance to this object is thought to be between 
$6.7$ and $9.1$ kpc \citep{papp08}, which could lead to a factor $\sim 2$ adjustment in the calculated X-ray luminosity in either case.

\begin{figure}
\includegraphics[width=\columnwidth]{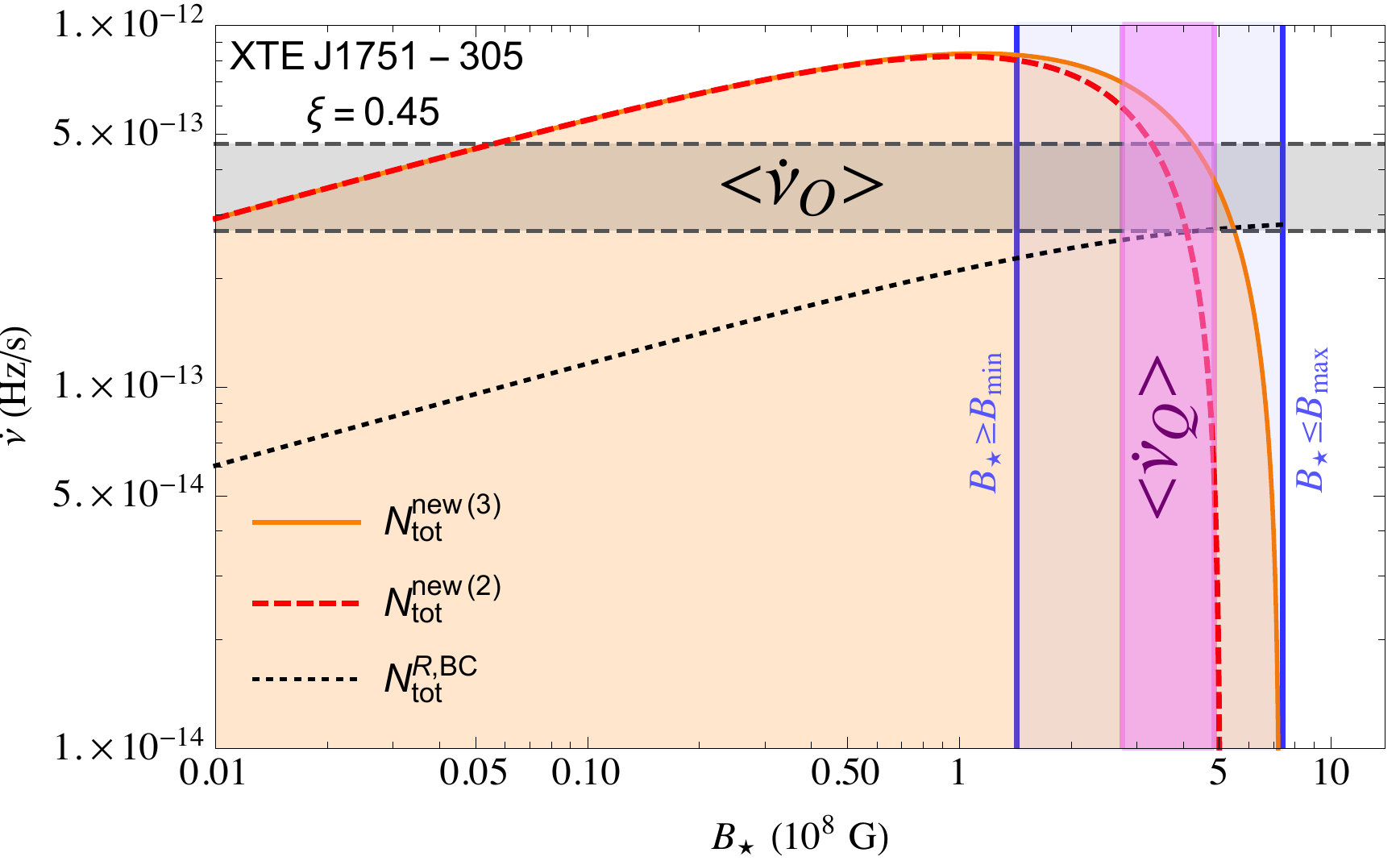}
\caption{A comparison between predicted spin-evolutions [via Eq.~\eqref{eq:torquebal}] for $N^{\text{R,BC}}_{\text{tot}}$ (dotted black curve), $N^{\text{new(2)}}_{\text{tot}}$ (dashed red curve),
and $N^{\text{new(3)}}_{\text{tot}}$ (orange curve), as functions of $B_{\star}$, for the 2002 outburst of J1751. The measured spin-up is shown 
in the grey band, while the blue lines to the left and right illustrate the minimum and maximum magnetic field strengths [for new(3)] permitted through the 
requirements $\RA > \Rs$ and $\RA < x_{\rm A, eq} \Rco$, respectively; see Eqs.~\eqref{Blow}-\eqref{Bhigh}. 
The region bounded by the magenta column represents the \emph{total} uncertainty in the magnetic field strength,
as calculated by combining the observational uncertainty in the mean spin-down $\langle \nu_{\text{Q}} \rangle$ and the variation 
$0 \leq \vartheta \leq \pi/2$ in the magnetic axis inclination.
 A fixed value of $\xi = 0.45$ is taken.}
\label{fig:xte}
\end{figure}

Figure \ref{fig:xte} compares the theoretical spin-up of J1751 during the 2002 outburst for the \cite{rap04} and \cite{bc17} model $N^{\text{R,BC}}_{\text{tot}}$ 
(dotted curve) with the new models $N^{\text{new(2)}}_{\text{tot}}$ (red) and $N^{\text{new(3)}}_{\text{tot}}$ (orange) considered here. In particular, since the 
former torque is the largest of all others [with the exception of new(2,3)] for the whole range of $x_{\text{A}}$ (see Fig. \ref{fig:torques}), it represents a maximum 
amongst the `classical' models. 

We see that the spin-up using $N^{\text{R,BC}}_{\text{tot}}$ is only marginally consistent with the observations: for the maximum predicted values of 
$\Bs$ from spin-down (see Tab. \ref{tab:nsdata}), the theoretical torque just scrapes the lower-limit of $\langle \nuO \rangle$. Since this torque is larger than 
the baseline models $N_{\text{A}}$ and $N_{\text{m}}$, our finding is consistent with those of \cite{ajh14}, since in this case the theoretical maxima would lie 
below the observed minimum. The situation is worse if one instead uses the smaller of the two X-ray luminosities discussed above, as was done by \cite{ajh14}. 
However, we see that both $N^{\text{new(2,3)}}_{\text{tot}}$ can comfortably accommodate even the maximum values of spin-up for $\xi = 0.45$ for the measured 
range of $B_{\star}$. Since values $\xi < 1$ are consistent with the prediction \eqref{xinum} for canonical values of disc thickness and viscosity, we conclude 
that the simple, analytic models considered here are consistent with the observations for this object. Note that the equilibrium values of the fastness parameter 
differ between models (2) and (3), and $N^{\text{new(2)}}_{\text{tot}}$ falls to zero at a lower $B_{\star}$ value than its slightly larger counterpart for $\xi = 0.45$ 
[see Eqs.~\eqref{xAeqn3} and \eqref{xAeqn2}]. Either way, we have that $B_{\rm max}$ exceeds the maximum allowed by spin-down.

\begin{figure}
\begin{center}
\includegraphics[width=0.8\columnwidth]{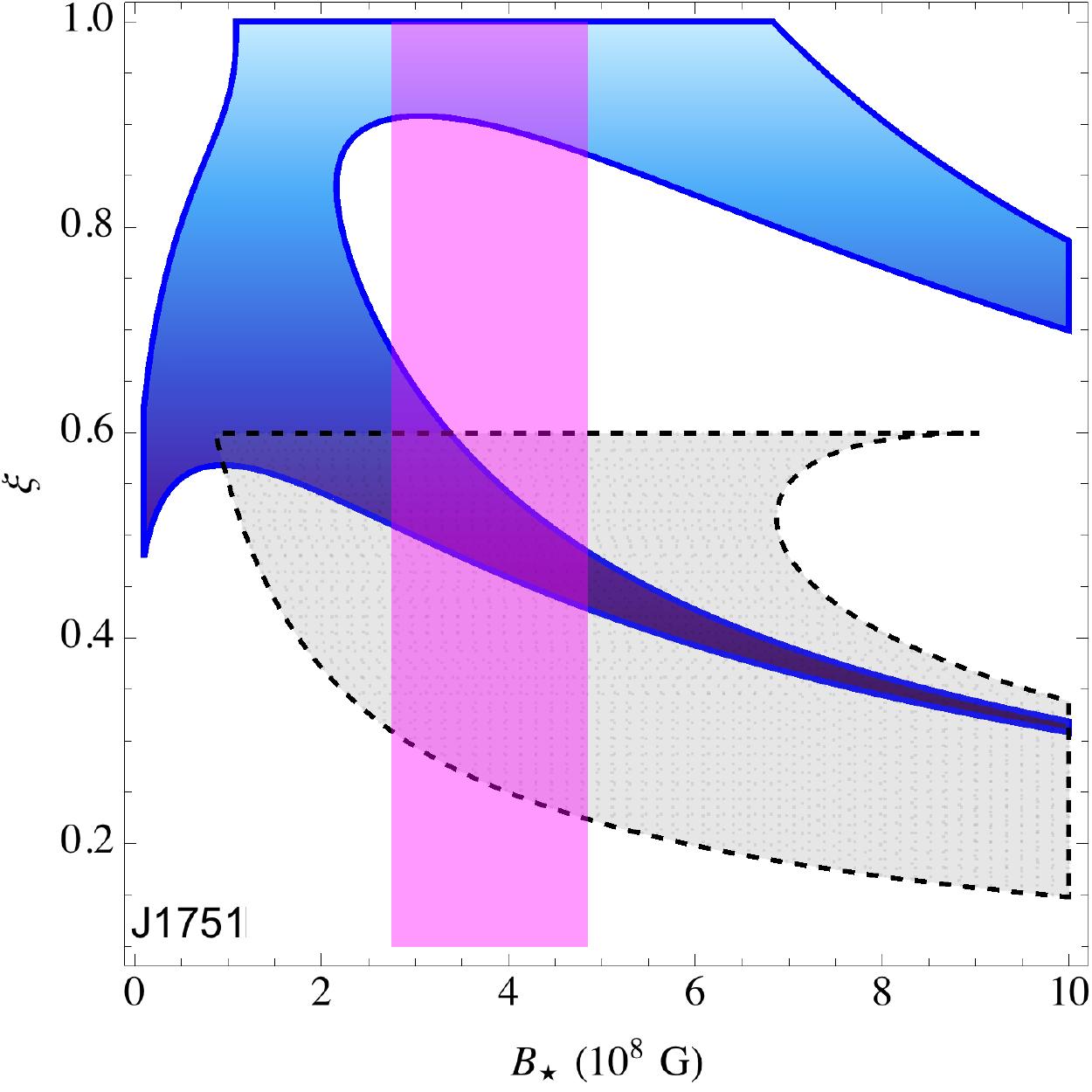}
\caption{The allowed $\xi-B_{\star}$ parameter space for the 2002 outburst of J1751. The blue region shows the theoretical combinations of $\xi$ and $B_{\star}$ for which $\dot{\nu}$, from $N^{\text{new(3)}}_{\text{tot}}$, takes a value within the range set by the outburst data. The magenta column shows the predicted range of $B_{\star}$ from quiescent spin-down, while the grey surface delimits the space over which $\xi$ and $B_{\star}$ respect the geometric requirements of the disc. The intersection between all three regions gives the range of $\xi$ and $B_{\star}$ consistent with observation for the torque model $N^{\text{new(3)}}_{\text{tot}}$.}
\label{fig:xte_xi}
\end{center}
\end{figure}

Although we take a fixed value of $\xi = 0.45$ above, there typically is, for a given torque model and $B_{\star}$, a range of $\xi$ such that the predicted spin-up lies within a desired band. In Figure \ref{fig:xte_xi} we show (blue curves) the theoretical $\xi-B_{\star}$ parameter space, associated with $N^{\text{new(3)}}_{\text{tot}}$, such that $2.7 \leq \dot{\nu}/(10^{-13} \text{ Hz s}^{-1}) \leq 4.7$ (see Tab. \ref{tab:nsdata}). The grey region instead shows the range of $B_{\star}$ and $\xi$ such that inequalities \eqref{Blow} and \eqref{Bhigh}, set by the geometric requirements of the disc, are satisfied. In particular, if $\xi$ is too large for some fixed $B_{\star}$ then the model does not permit an equilibrium fastness parameter [see also Eq.~\eqref{eq:eqfast}], while if $\xi$ is too small for some fixed $B_{\star}$ then the Alfv{\'e}n radius will cut into with the star. The final region of interest in Fig. \ref{fig:xte_xi} is the magenta column which shows, as in Fig. \ref{fig:xte}, the $B_{\star}$ range predicted by spin-down. The complicated shape that is formed by the intersection of all three surfaces described above yields the theoretically- and observationally-allowed parameter space for this object. Note, however, that this range is itself sensitive to the other (uncertain) parameters intrinsic to the system (e.g., $M_{\star}$, $R_{\star}$, $\dot{M}$, ...).

%

\subsection{IGR J00291+5934}
\label{sec:IGR}

Table \ref{tab:nsdata} reports data relevant for the 2004 and 2015 outbursts of J00291, which are discussed in detail below. Some notes are as follows.
Assuming the 2015 burst was of a pure helium nature, \cite{defalc17} constrained the distance of J00291 to be $4.2 \pm 0.5$ kpc. \cite{falanga05} 
report a \emph{peak} luminosity of $6.3 \times 10^{36} \text{ erg s}^{-1}$, though assumed $d = 5$ kpc [see also \cite{ajh14}]. The luminosity for the 
2015 outburst is inferred from data given in \cite{tudor17}, who did not include comptonization or bolometric corrections, by adopting the correction 
factor used by \cite{falanga05} for the 2004 burst. Note however that the peak and mean values for both the 2004 and 2015 bursts differ between 
\cite{defalc17} and \cite{falanga05,tudor17}, respectively: the former authors suggest a higher $(\sim 40\%)$ flux (see Table 2 therein). 

\begin{figure}
\includegraphics[width=\columnwidth]{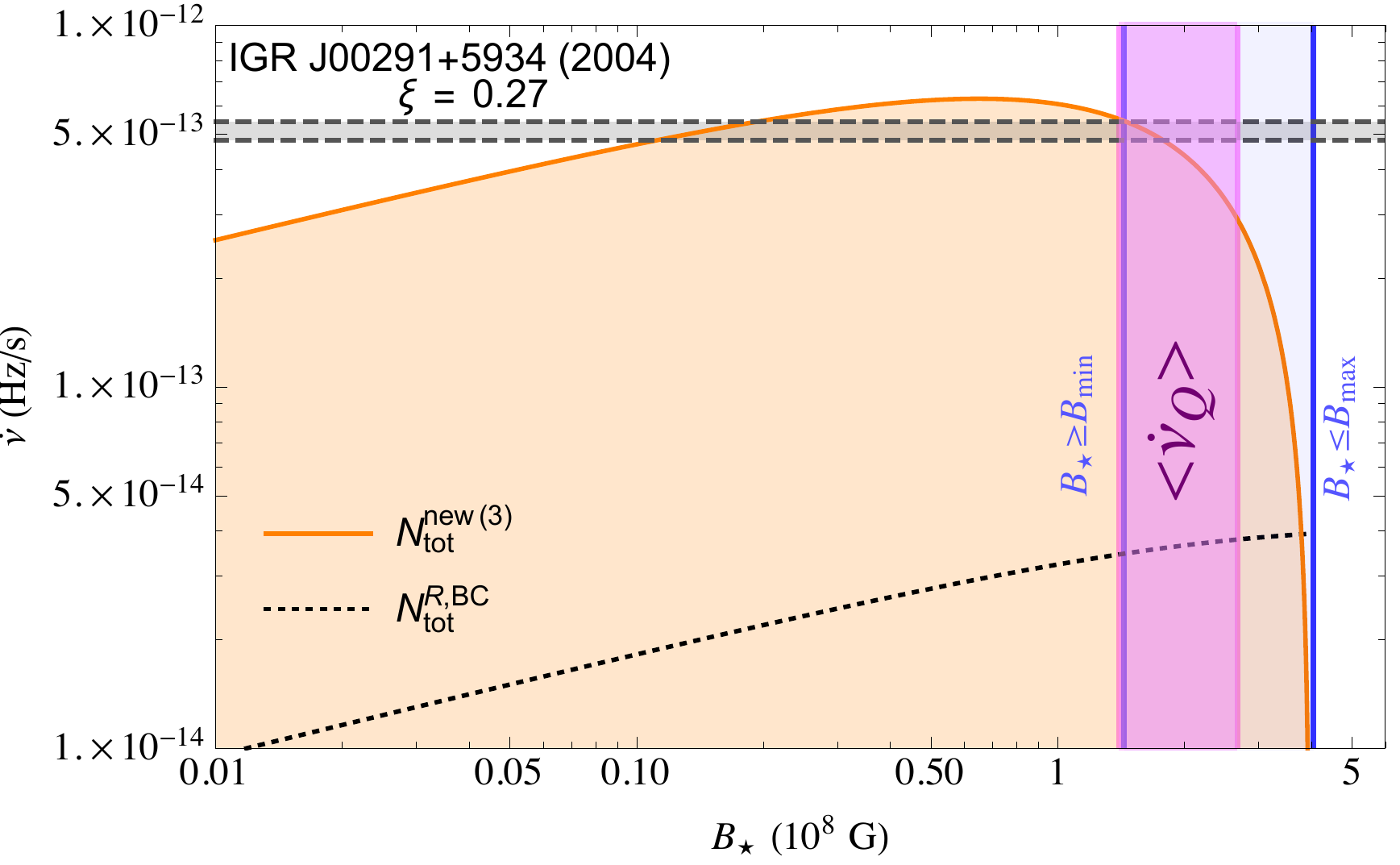}
\caption{Similar to Fig. \ref{fig:xte} but for the 2004 outburst of J00291. A fixed value of $\xi = 0.27$ is taken.}
\label{fig:igr2004}
\end{figure}

Similar to Fig. \ref{fig:xte}, Figure \ref{fig:igr2004} compares theory with observation for the 2004 outburst of J00291. In this instance, again noting that 
$N^{\text{R,BC}}_{\text{tot}}$ is the largest amongst the `classical' models, we see that the various (analytic) torque expressions thus far considered 
in the literature are utterly unable to accommodate the spin-up for this object, as concluded by \cite{ajh14}. In particular, even for extreme values of 
$\Bs \lesssim 5 \times 10^{8} \text{ G}$ -- the maximum allowed by the requirement that $\RA < x_{\rm A, eq} R_{\text{co}}$ 
-- the largest value of $\dot{\nu}$ from 
$N^{\text{R,BC}}_{\text{tot}}$ is an order of magnitude below the reported value $\langle \nuO \rangle$. Furthermore, tweaking the mass, radius, 
or moment of inertia of the star within reasonable ranges is not able to alleviate the discrepancy. By contrast, for the model $N^{\text{new(3)}}_{\text{tot}}$ 
with $\xi = 0.27$, we see that the whole range of the observed spin-up can be met. Similar conclusions are found for $N^{\text{new(2)}}_{\text{tot}}$.

\begin{figure}
\includegraphics[width=\columnwidth]{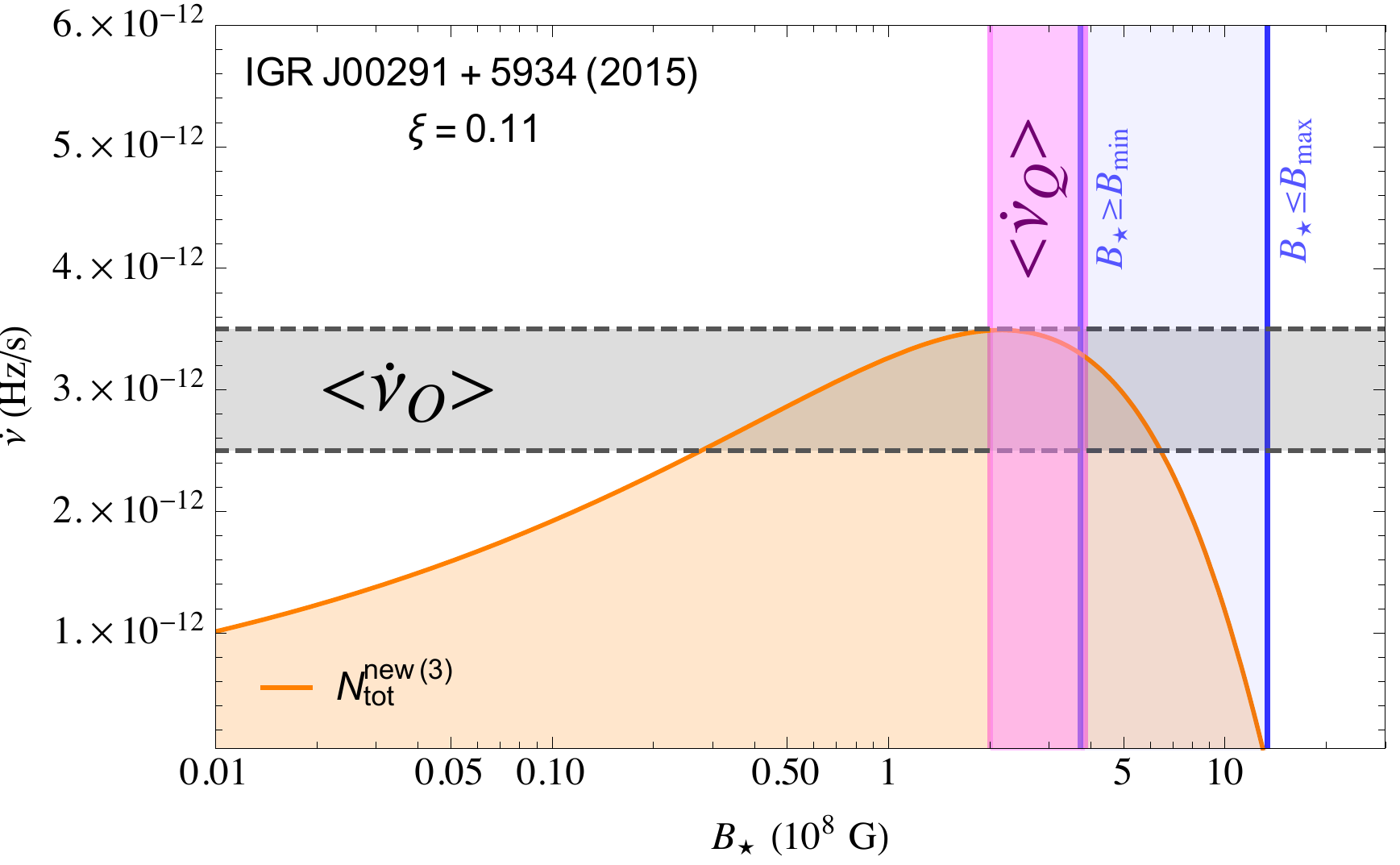}
\caption{Theoretical spin-up predicted using $N^{\text{new(3)}}_{\text{tot}}$ (orange curve), as a function of $B_{\star}$, 
for the 2015 outburst of J00291. Other torque models are not shown, since they lie well-below the y-axis range shown here. 
A fixed value of $\xi = 0.11$ is taken.}
\label{fig:igr2015}
\end{figure}

In 2015 the object went into outburst again, though this time displayed a $\gtrsim 6$ times higher degree of spin-up than in 2004 \citep{sanna17b}, even 
though the X-ray luminosity was smaller by a factor $\lesssim 4$ [see Tab. \ref{tab:accmodels}; though cf. \cite{defalc17}]. As such, unless we seriously 
underestimate the accretion rate $\dot{M}$ (cf. Sec. \ref{sec:LT}), the `classical' models fall very short (by $\sim 2$ orders of magnitude) of being able to 
explain the spin-up in this case. If instead we take a value $\xi = 0.11$ and a slightly more compact star than in Fig. \ref{fig:igr2004} with 
$G \Ms /c^2 \Rs \sim 0.24$ however, even the extreme, upper-limit value of $\langle \nuO \rangle$ can be matched using 
the torque $N^{\text{new(3)}}_{\text{tot}}$, as shown in Fig. \ref{fig:igr2015}. This matching however requires the magnetic field to be on the 
low end within the allowed region, i.e., that $\RA \sim R_{\star}$. GR effects, which we have thus far ignored, may therefore be important since 
the spacetime is expected to be strongly non-Minkowski near the stellar surface (see Sec. \ref{sec:GRcorrect}). 

\begin{figure}
\begin{center}
\includegraphics[width=0.49\columnwidth]{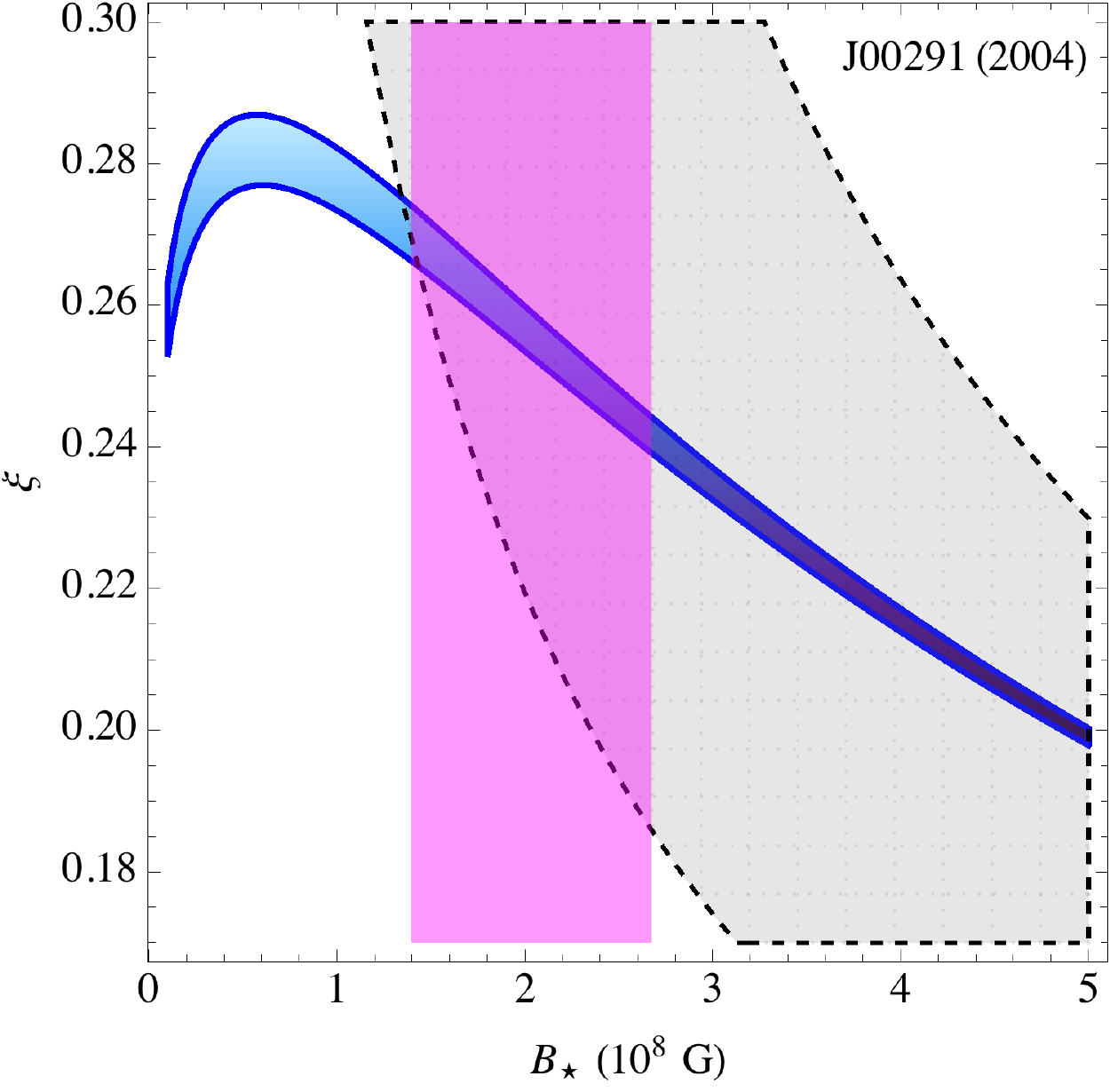}
\includegraphics[width=0.49\columnwidth]{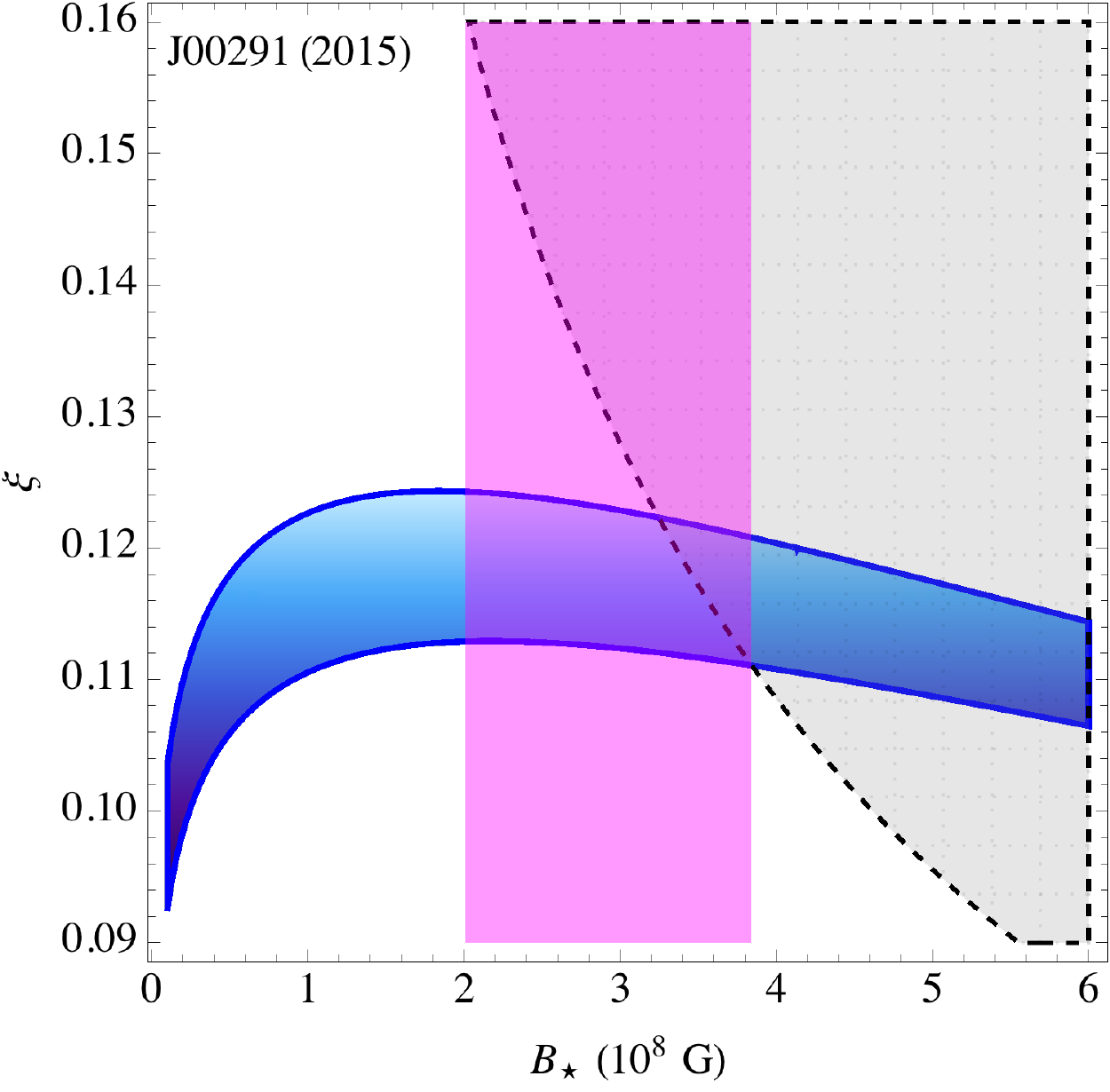}
\caption{Similar to Fig. \ref{fig:xte_xi}, though for the 2004 (left panel) and 2015 (right panel) outbursts of J00291. Note that slightly different compactness values are taken between the two cases for demonstration purposes, so that the spin-down constraints on $B_{\star}$ (magenta columns) differ slightly.}
\label{fig:igr_xi}
\end{center}
\end{figure}

Similar to Fig. \ref{fig:xte_xi}, Figure \ref{fig:igr_xi} shows allowed combinations of $\xi$ and $B_{\star}$ for the 2004 (left panel) and 2015 (right panel) outbursts of J00291. The blue strips show the theoretical values of $\dot{\nu}$, again calculated from $N^{\text{new(3)}}_{\text{tot}}$, consistent with the observed spin-ups, while the grey region delimits the geometrically-set ranges of $\xi$ and $B_{\star}$. Note also that since we take slightly different compactness values between Figs. \ref{fig:igr2004} and \ref{fig:igr2015} for demonstration purposes, the observational range of $B_{\star}$ [Eq.~\eqref{eq:dipnu}] and the theoretical range of $\xi$ and $B_{\star}$ [Eqs.~\eqref{Blow} and \eqref{Bhigh}] differs between the two cases. We see that in either case the constraints on $\xi$ and $B_{\star}$ are stricter than for J1751; for the 2004 data this is because the error bars on $\langle \dot{\nu}_{O} \rangle$ are much tighter, while in 2015 the spin-up was so extreme that it is difficult to produce the required torque unless $\xi$ is taken close to the theoretical minimum where $R_{A} \gtrsim R_{\star}$, as described above. Physically speaking, these findings suggest that the disc in this system may be less viscous and/or thinner than for J1751, and that $\alpha$, $H$, or $R_{\text{A}}$ are dynamical over $\sim$ year-long timescales; see expression \eqref{xinum}.

%

\subsection{SAX J1808.4--3658}
\label{sec:SAX}

Finally, we also consider the system J1808, which was not studied by \cite{ajh14}. Some relevant notes are as follows. The quiescent spin-down rate from \cite{hart09} [$\langle \nuQ \rangle = -5.5(12) \times 10^{-16} \text{ Hz s}^{-1}$] differs from \cite{sanna17} [$\langle \nuQ \rangle = -1.5(2) \times 10^{-15} \text{ Hz s}^{-1}$], possibly because spin-down was accelerated during those 8 years or because spin-up during outburst episodes were handelled in a statistically different way [see \cite{sanna17} for a discussion]. More extremely, for the 2002 burst data specifically, \cite{burd06} report a value of $\langle \nuQ \rangle = -7.6(15) \times 10^{-14} \text{ Hz s}^{-1}$ for the spin-down. Using the traditional braking formula \eqref{eq:dipnu} thus implies a factor $\sim 10$ larger dipole moment relative to the other epochs, because of the factor $\sim 100$ increase in the magnitude of $\dot{\nu}_{\text{Q}}$ (see fourth column of Tab. \ref{tab:nsdata}). \cite{burd06} suggest that a smaller value of $B_{\star} \lesssim 4 \times 10^{8} \text{ G}$ is obtained if one instead assumes that the spin-down torque is due to the magnetic drag on the accretion disc [specifically, they apply Eq. (23) of \cite{rap04}]. Taken literally in the context of expression \eqref{eq:dipnu} however, these timing fits suggest that the spin-down of the object varies substantially on timescales of $\sim 10$ yr, possibly indicating rapid magnetic field evolution or the formation of an accretion-built mountain. We note that distance measurements for this object are rather tight, viz. $d = 3.5(1)$ kpc \citep{gal06}, so much larger values of $L_{\rm X}$ are unlikely. Data for the 2015 outburst come from the XMM Newton measurements rather than those from NuSTAR, the latter of which predicts an even more extreme spin-up (by a factor $\sim 10$). \cite{sanna17} suggest that the NuSTAR measurements are overly large because of the time drift within the internal clock of the instrument.

\begin{figure}
\includegraphics[width=\columnwidth]{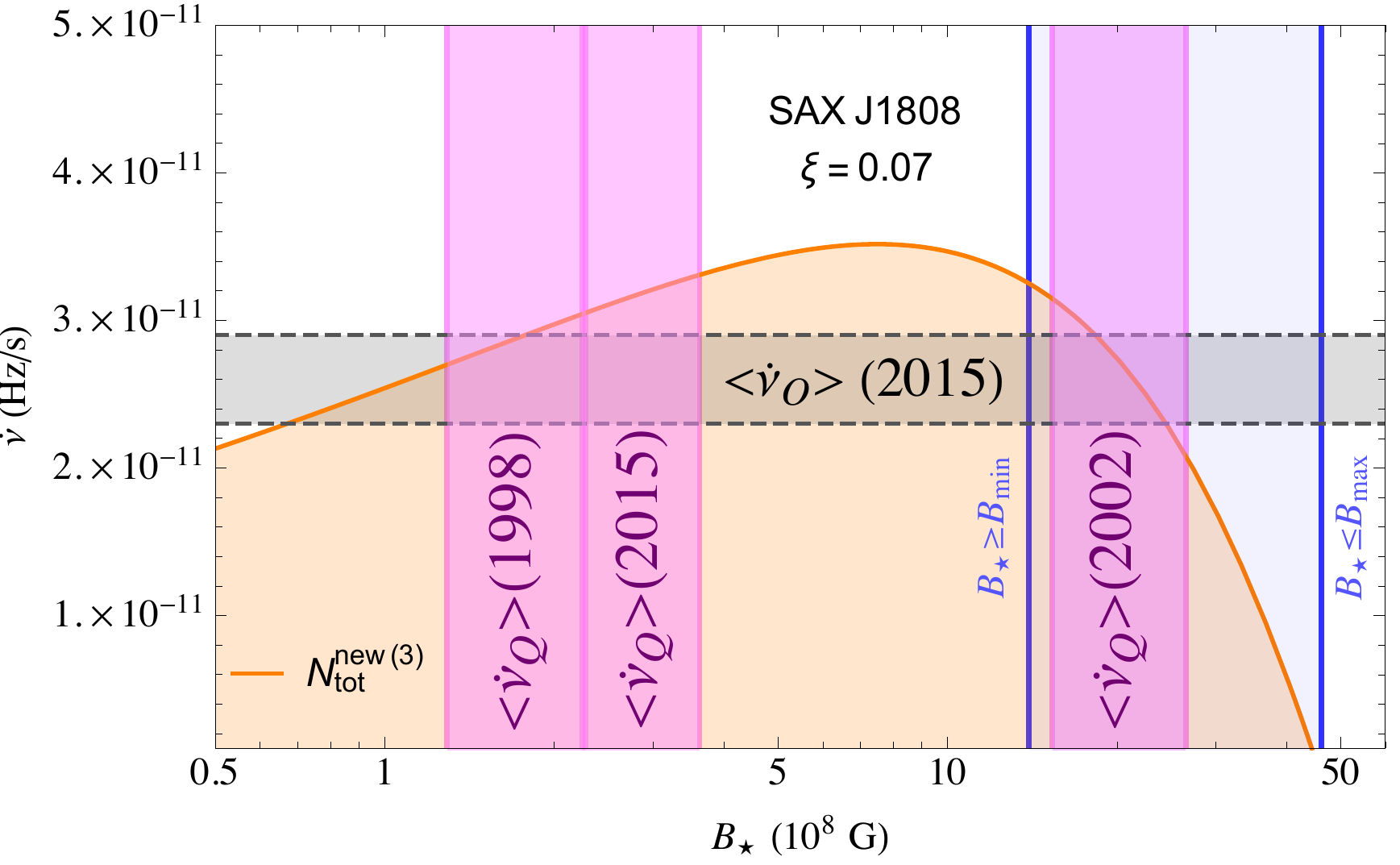}
\caption{Theoretical spin-up predicted using $N^{\text{new(3)}}_{\text{tot}}$ (orange curve), as a function of $B_{\star}$, for the 2015 outburst of J1808. 
For this object we take a compact star with $\Ms = 2.0 M_{\odot}$ and $\Rs = 10$ km, in line with the findings of \protect\cite{li99}. Other torque models 
are not shown, since they lie well-below the y-axis range shown here.  We show the predicted $B_{\star}$ values from spin-downs reported by different 
authors for separate epochs in the various magenta columns (see Tab. \ref{tab:nsdata}). A fixed value of $\xi = 0.07$ is taken.}
\label{fig:sax}
\end{figure}

Figure \ref{fig:sax} compares the theoretical spin-up predicted by $N^{\text{new(3)}}_{\text{tot}}$ for the 2015 outburst of J1808. In particular, the spin-up achieved during this period, according to \cite{sanna17}, is the most extreme of all systems thus far observed. The X-ray luminosity found during the outburst was not particularly high however, and so, much like in the case of J00291, the `classical' models are not able to come close to explaining the spin-up here (though they can for the 1998 and 2002 outbursts). In fact, the spin-up is so large that even the model $N^{\text{new(3)}}_{\text{tot}}$ cannot account for the data unless we use the spin-down estimates for the 2002 burst \citep{burd06} and take $\xi \lesssim 0.1$. In particular, using the spin-down values $\langle \nuQ \rangle$ for either the 1998 and 2015 cases returns $B_{\star}$ values which are much 
smaller than the requirement set by $\RA > R_{\star}$, see Eq.~\eqref{Blow}.

It is interesting to note however that in 2008, XMM-Newton and Suzaku captured a relativistically-broadened K-$\alpha$ iron line at $\sim 6.5$ keV in the spectrum of J1808 \citep{papp09,cack09}. It is generally thought that these emission lines originate from the inner edge of the accretion disc, and therefore their spectra can, in principle, be used to determine the magnetospheric radius \citep{amxp_review}. For J1808, the emission spectra suggest this radius lies at $\sim 4.4^{+1.8}_{-1.4}$ Schwarzschild radii \citep{papp09}. For a star with $M = 2.0 M_{\odot}$, as found by \cite{li99}, we therefore obtain $B_{\star} \gtrsim 2 (\xi/0.1)^{-7/4} \times 10^{9} \text{ G}$ by matching expression \eqref{RA} with the above, which agrees with the value inferred from spin-down measured in 2002 and the values needed to explain the 2015 outburst.

Regardless, there appears to be some conflict between the ranges of $B_{\star}$ inferred from different epochs. We are therefore left with a few possible conclusions. (i) The reported spin-up in 2015 is an overestimate, possibly for the reasons described in Sec. \ref{sec:obs} or in \cite{sanna17}. (ii) 
The reported spin-down (for the 2015 burst) is too low, or (iii) there is some physics that becomes important for extreme values of $\xi \lesssim 0.1$ that is not 
included in the description of the torque models. To conclusively rule out option (iii) one requires 3D simulations of realistic matter flows in accreting neutron 
stars, such as those described in \cite{kulk13}. In any case, the message here is that there are still several unanswered questions concerning spin evolution in AMXPs.

\section{Additional physics to consider}
\label{sec:morephysics}

Accreting neutron stars are almost by default rather `dirty' physical systems and as a consequence
the magnetospheric accretion model described in Section~\ref{sec:models} is unlikely to capture 
all of the relevant physics. In this section we discuss in more detail some of the 
most important corrections to that model, namely, the likely multipolar structure of the magnetic field and the 
impact of GR gravity. 
It should be pointed out that none of these effects change the main conclusions of this paper but could help alleviate 
the tension seen in some systems as in, for example, J00291's spin-up during its 2015 outburst (see Fig.~\ref{fig:igr2015}).

\subsection{Multipolar magnetic fields}
\label{sec:multiB}

Assuming a force-free magnetosphere\footnote{Though such a description may not be valid in the accretion layer near 
the stellar surface, where diamagnetic screening currents reside \citep{choud02}, it is likely a fair description at the Alfv{\'e}n radius 
\citep[see, e.g., the Grad-Shafranov simulations of][]{wet10,suvm20}.} the poloidal field can be expressed as a sum of force-free multipoles. 
In most models considered in the literature, only the dipole component is kept for simplicity, as in Sec. \ref{sec:Models}. However, recent observations of hot 
spot activity on PSR J0030+0451 \citep{bilous19} and GRO J1744--28 \citep{doro20}, together with cyclotron resonant scattering features 
seen in several accretion-powered X-ray pulsars \citep{stau19}, suggest that the magnetic fields of stars with a history of accretion 
are likely to contain non-negligible multipole components near the stellar surface. Theoretical considerations support this conclusion, as comparable multipole 
moments are seen to form from seeding dipole fields in simulations of crustal Hall drift coupled with (accretion-accelerated) Ohmic decay 
\citep{gep95,rhein02,cumm04} and accretion-induced magnetic burial \citep{pri11,suvm20}. Nevertheless, since a general $\ell$-pole falls 
off like $r^{-(2+1) \ell}$, the dipole component will typically dominate at large radii $(r \gg \Rs)$. Unless the field is sufficiently weak therefore 
such that the Alfv{\'e}n radius lies close to the stellar surface, one can typically ignore higher-multipoles for the purpose of accretion torque 
modelling. For completeness however, we consider here a field with a strong quadrupole to illustrate the impact of multipolar components 
on the behaviour of the accretion torque.

\begin{figure}
\includegraphics[width=1.0\columnwidth]{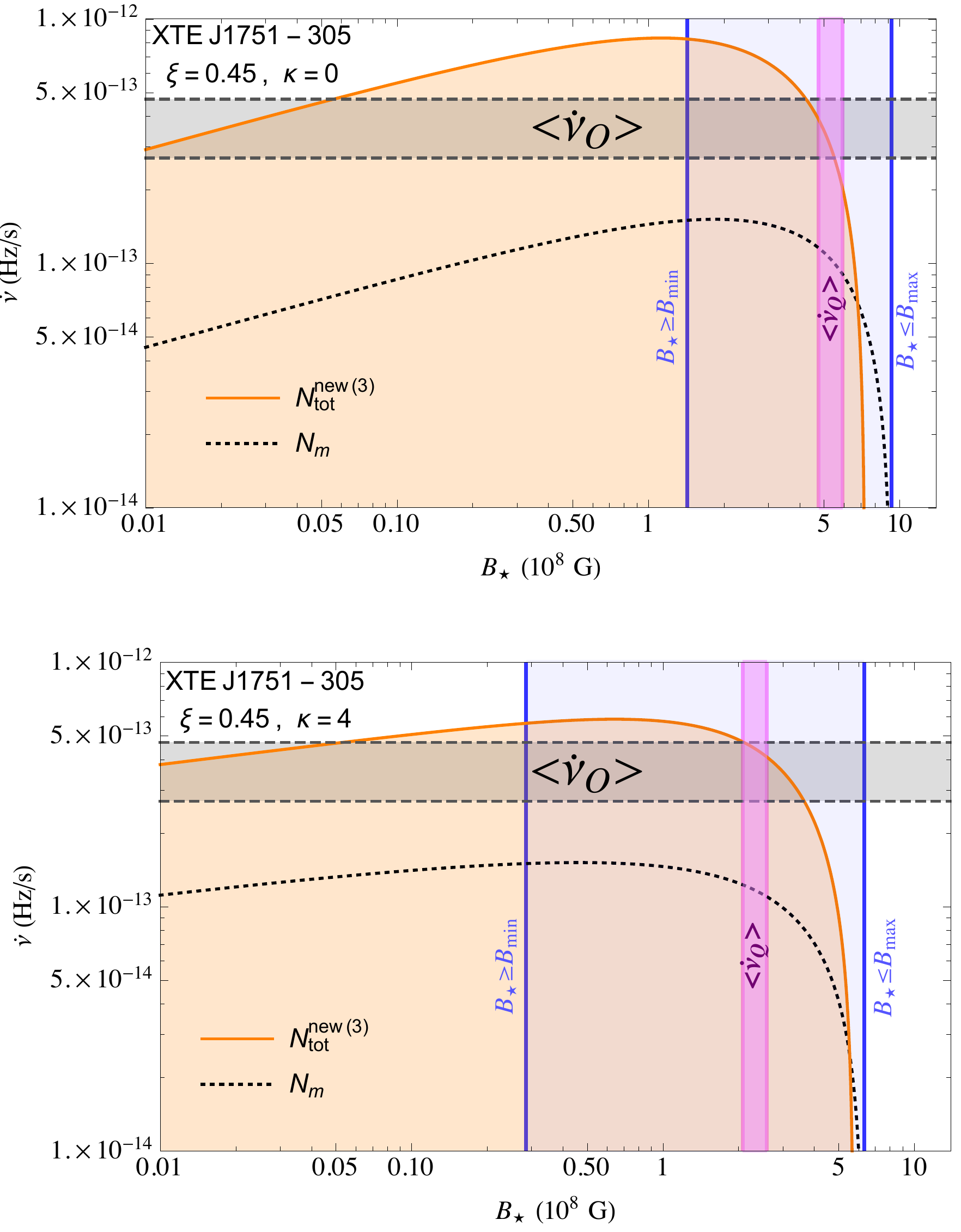}
\caption{Baseline magnetospheric (black, dotted) and new(3) (orange, solid) accretion torques, as functions of the characteristic magnetic 
field strength $B_{\star}$, for J1751. The top panel shows the pure dipole case $(\kappa = 0)$, while the bottom panel illustrates a 
case with a strong quadrupole component, $\kappa = 4$ [see Eq.~\eqref{eq:dipquadreln}]. Each torque assumes canonical stellar parameters, 
$\xi = 0.45$, and the spin-down formula \eqref{eq:dipquadnu} to determine the range of $B_{\star}$ values that agree with the quiescent measurement 
$\langle \nuQ \rangle$. Note that $B_{\text{max}}$ corresponds to the torque $N_{\text{m}}$.
}
\label{fig:dipquad}
\end{figure}

This is achieved by introducing a dimensionless parameter $\kappa$, which quantifies the strength of the quadrupole field, through
\be
\label{eq:dipquadreln}
B_{z}  = - \frac{\Bs \Rs^3}{r^3} \left( 1+ \kappa \frac{\Rs^2} {r^2} \right).
\ee
Non-dipolar terms directly influence the Alfv{\'e}n radius, as the roots of the Euler equation \eqref{EulerMHD} are 
necessarily shifted (see also Sec. \ref{sec:multiB}). For magnetic fields consisting of mixed multipoles, finding these roots generally requires numerical methods.
The functional forms for the accretion torques, given as integrals over some weighted magnetic energy density, are also adjusted. 
For instance, in the spirit of `mechanism (3)' described by \cite{wang95}, we find, from Eq.~\eqref{Btor},
\begin{align}
N_{\text{disc}} &= -\int^{\infty}_{\RA} dr r^2 f_{(3)}(r) B_{z}(r)^2 
\nn \\
&= N_{\text{dip}} \Bigg[ 1 + \frac{18 \kappa}{455} \left ( \frac{\Rs}{\RA} \right )^2
\left ( \frac{91 - 130 \oA + 18 \oA^{10/3}} {3 + 2 \oA^2 - 6 \oA} \right ) 
\nn  \\
 & \,\,\,\,\,\, + \frac{9 \kappa^2}{1309}  \left ( \frac{\Rs}{\RA} \right )^4 \left (  \frac {187 - 238 \oA + 18 \oA^{14/3}} 
 {3 + 2 \oA^2 - 6 \oA} \right ) \Bigg].
\label{eq:quadtor}
\end{align}
Using expression \eqref{eq:quadtor}, we compare the total torques, $N_{\text{A}} + N_{\text{disc}}$, obtained for pure dipole $(\kappa = 0)$ and strong 
quadrupole $(\kappa = 4)$ cases, for parameters relevant to J1751, in Fig.~\ref{fig:dipquad}. Including a quadrupole component tends to flatten the 
torque curve, i.e., $\Ntot^{\text{new(3)}}$ varies slower as a function of $\Bs$ for greater $\kappa$. For fixed values of $\xi$, we see that 
the maximum torque that can be achieved is lower in the quadrupole case (by a factor $\sim 2$ for $\kappa \sim 4$) though, for $\xi = 0.45$, 
the mixed case is still able to accommodate the upper limits set by \cite{papp08} for J1751's spin-up. We see also that $\Nm$, 
shown by the dotted curve, lies well below the spin-up band with or without quadrupole fields. Including a quadrupole component shifts the minimum 
and maximum values of the $B$ field (as detailed in Sec. \ref{sec:Bcons}), as set by the geometrical requirements of the magnetosphere, 
to the left. Specifically, since the quadrupole is strong at the stellar surface, the spin-down and minimum are shifted more noticeably than the maximum, 
which is set by the physics occurring near the co-rotation radius. Note in particular that the electromagnetically-induced spin-down for a mixed 
dipole-quadrupole field reads \citep{petri19}
\be
\dot{\nu}_{\rm Q} = - \frac {2  \pi^2 \mus^2  \nuS^3} {3 \Is c^3} \left( 1 + \kappa \right)
 \left [ \, 1 +  \frac{64  \pi^2}{45} \kappa^2 \left ( \frac{ \Rs \nuS}{c} \right )^2 \,\right ] ,
\label{eq:dipquadnu}
\ee
where in the absence of a Spitkovksy-like formula for the field considered here we have assumed the standard model of an orthogonal rotator 
in vacuum (i.e., $K = 1/3$). This expression implies that the inferred $\Bs$ is also sensitive to $\kappa$, as can be seen from the magenta bars 
in Fig. \ref{fig:dipquad}.

\subsection{General-relativistic corrections}
\label{sec:GRcorrect}

Throughout our analysis thus far, we have restricted our attention to Newtonian equations of motion. There are, however, a number of places 
where GR corrections are likely to play a role, especially for very compact stars [such as J1808 \citep{li99}]. In GR, the Euler and induction
equations [Eqs.~\eqref{EulerMHD} and \eqref{induction}] become weighted by the spacetime metric coefficients, most notably by the `$tt$' Schwarzschild 
redshift factor, $z \sim 1 - 2 G \Ms / c^2 \Rs$. For rapidly rotating stars [such as J00291], rotational corrections to the geometry, including precession 
(see Sec. \ref{sec:LT}), may also become important as Birkhoff's theorem can no longer be faithfully applied to describe the spacetime exterior to the star 
\citep[see, e.g., ][]{papp12}. These factors will shift the geometric radii important to the accretion problem, such as $\RA$ and $\Rco$. Ultimately however, 
many of these corrections can be absorbed by our phenomenological parameters, such as $\xi$ [i.e., by making the replacement 
$\xi_{\text{GR}} \approx z^{-1} \xi_{\text{N}}$], which depend sensitively on the flow particulars and are highly uncertain. Including relativistic corrections 
self-consistently requires one to solve the full GR-MHD system of equations, which is beyond the scope of this work [cf. \cite{kulk13}].

One can however get a clean but rough estimate for the importance of GR corrections by considering Post-Newtonian (PN) expansions. 
That is, by expanding the Einstein equations in powers of $c^{-2}$. As shown by \cite{blanc98}, the Keplerian velocity profile at 2PN reads
\be
\label{eq:2pnomega}
\OmK^{2\text{PN}}(r) = \OmK (r) \left( 1 - 3 \cC \frac{\Rs} { r} +  6 \cC^2 \frac{\Rs^2} {r^2} \right)^{1/2},
\ee
where
\be
\cC = \frac{G \Ms}{c^2 \Rs},
\label{compact}
\ee
is the stellar compactness. 
Though somewhat tedious, one could simply repeat the calculations performed in Sec. \ref{sec:Models} using the rotational profile \eqref{eq:2pnomega} 
instead of the standard Keplerian one, $\OmK$. Consider just Eq.~\eqref{EulerMHD}, which implies that the Alfv{\'e}n radius resides at the solution to
\be 
\label{eq:2pnRA}
\dot{M} \frac {d} {dr} \left[ \Omega_{\text{K}}^{2\text{PN}}(r) r^2 \right]_{\RA} = - \RA^2 \left( B_{\phi} B_{z} \right)_{\RA}.
\ee
Even for a dipolar magnetic field, the roots of expression \eqref{eq:2pnRA} must be found numerically because of the high-order nature of the 
polynomial involved. Similarly, the corotation radius, defined as the point where $\Omega_{\text{K}}^{2\text{PN}}(\Rco) = 2 \pi \nuS$, must also be 
evaluated numerically. 

\begin{figure}
\includegraphics[width=\columnwidth]{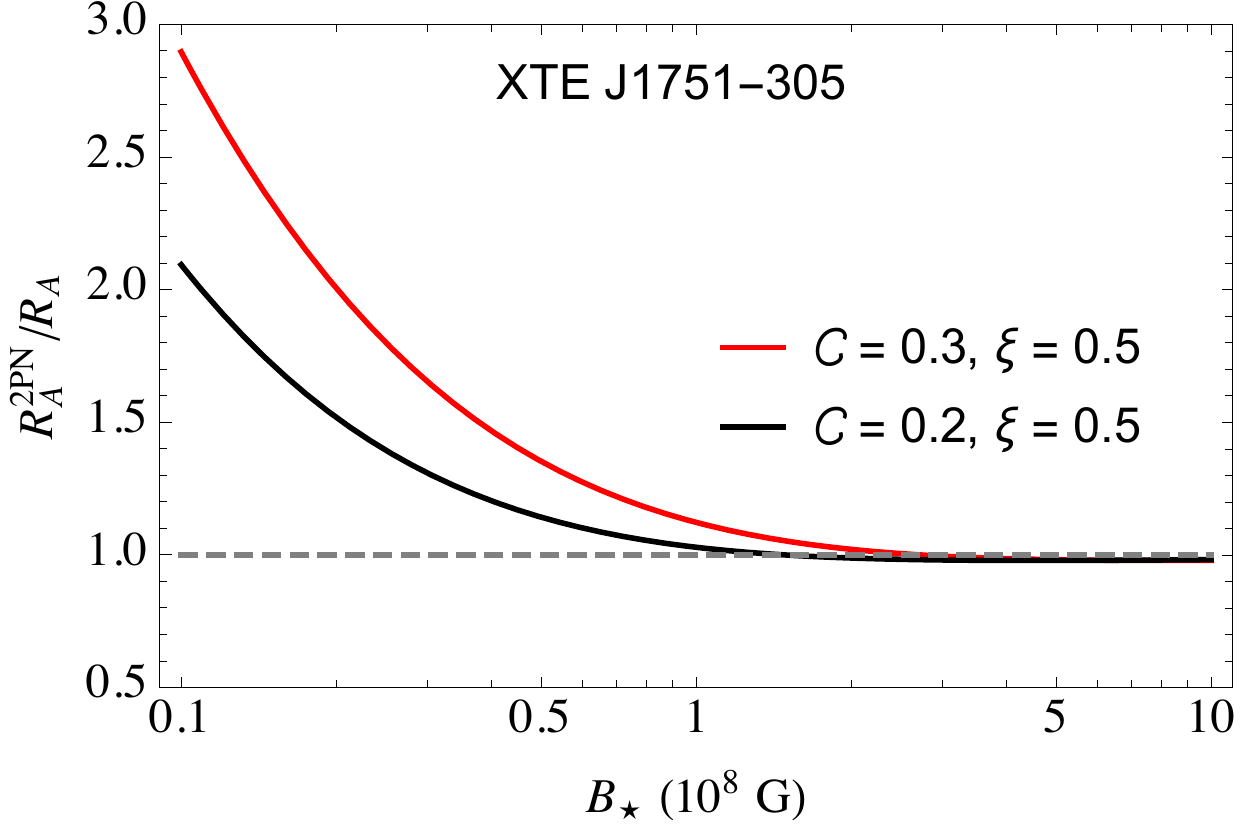}
\caption{Normalised 2PN Alfv{\'e}n radii as functions of $B_{\star}$ for stars with compactness $\cC = 0.2$ (black curve) and 
$\cC = 0.3$ (red curve). A value of $\xi = 0.5$ is chosen, though smaller (larger) values lead to a greater (lesser) shift relative to the 
Newtonian value, $\RA$. We take an accretion rate $\dot{M} \sim 10^{-9} M_{\odot}/\text{yr}$, as appropriate for J1751 (see Tab. \ref{tab:nsdata}).}
\label{fig:2pn}
\end{figure}

Figure \ref{fig:2pn} shows two different Alfv{\'e}n radii at 2PN order as functions of the magnetic field strength $B_{\star}$ for J1751. 
In particular, we consider the case of a less compact star with $\cC= 0.2$ (black curve) and a more compact one with $\cC = 0.3$ (red curve). 
As expected, the deviation, relative to the Newtonian approximation \eqref{RA}, is larger for the more compact case ($\sim 20\%$ larger at 
$\Bs = 10^{8} \text{ G}$). As the magnetic field strength increases, the Alfv{\'e}n radius moves further away from the stellar surface, and the PN terms 
become less important. As such, in both examples with $\xi = 0.5$ (ignoring the distinction between $\xi_{\text{N}}$ and $\xi_{\text{GR}}$), the corrections 
become negligible for $\Bs \gtrsim 3 \times 10^{8} \text{ G}$ where $\RA \gtrsim 2 \Rs$. Note, however, that for smaller values of $\xi$ or larger values of 
$\cC$, the Alfv{\'e}n radius moves closer to the surface of the star [as can be seen from expression \eqref{RA}] and the PN terms remain important for a 
wider range of $\Bs$. For J1751, where a value $\Bs \approx 5.3 \times 10^{8}$ G is predicted from spin-down \citep{rigg11}, we conclude that PN corrections 
are likely to be negligible. However, since $\nuS = 435$ Hz for this object, we can also calculate that the corotation radius is shifted by a factor 
$\Rco^{\text{2PN}} / \Rco \approx 0.92$ for a compactness $\cC = 0.25$. Since this ratio is smaller than one, this implies that the fastness parameter 
$\oA$ is larger than its Newtonian counterpart (by $\sim 10\%$), which leads to a marginally smaller torque in most models (cf. Fig. \ref{fig:torques}).

\subsection{Disc tearing via Lense-Thirring precession}
\label{sec:LT}

For a rotating source, Lense-Thirring (LT) precession appears as soon as we move from Newtonian to GR gravity. 
This key effect should be present when the disc is locally misaligned with the stellar spin axis, in other words
when the disc is locally non-equatorial~\citep{bp75}. LT dynamics have been extensively studied in accreting black holes; 
detailed calculations suggest that it may cause the disc to fragment in a series of precessing rings~\citep{nixon12,raj21}. 
In this section we retrace the calculation of~\cite{nixon12}, although having LMXBs in mind instead of black holes. 

To leading PN order, the stellar spin-driven LT torque on the disc is given by the formula,
\be
N_{\rm LT} = 2\pi r H | \mathbf{\Omega}_{\rm LT} \times \mathbf{L} |,
\ee
where
\be
\mathbf{\Omega}_{\rm LT}  (r) = \frac{2G}{c^2 r^3} \mathbf{J}_\star,
\ee
is the vectorial LT angular precession frequency, $\mathbf{J}_\star$ is the stellar angular momentum
and $L (r) = r^2 \Sigma \Omega $ is the disc's angular momentum per unit area. We then have,
\be
N_{\rm LT} = 2\pi \sin\theta r^3 H \Sigma \Omega_{\rm LT} \Omega, 
\ee
where $\theta$ is the (local) angle between $\mathbf{J}_\star$ and $\mathbf{L}$. 

The LT precession is counteracted by viscosity in the disc. The associated viscous torque 
(along the spin axis) 
is,
\be
N_{\rm visc} = 2\pi \nu r^3   \Sigma\Omega^\p.
\ee
Hereafter the disc is taken to be Keplerian,  $\Omega = \OmK$.

The disc is likely to undergo tearing by the precessional motion provided the LT torque exceeds the viscous torque, 
\be
N_{\rm LT} \gtrsim N_{\rm visc}.
\label{LTcond1}
\ee
For a uniform density star $J_\star =  (2/5) \Ms \Rs^2 \Oms $;  after some straightforward algebra and with the 
help of the disc structure Eqs.~\eqref{discparams}, the above inequality leads to
\be
\left ( \frac{r}{\Rs} \right )^{3/2} \lesssim \frac{8}{15}\sin\theta  \frac{\cC}{\alpha} 
 \frac{\Oms}{\Omega_0}  \left (\frac{H}{r} \right )^{-1}, 
\ee
where we have introduced the `Kepler limit' frequency $ \Omega_0 = \OmK (\Rs) $
and the compactness $\cC$ is defined in Eq.~\eqref{compact}.  

Expressing our result in terms of normalised parameters, we have
\begin{align}
\frac{r}{\Rs}  &\lesssim 40 (\sin\theta)^{2/3}    \left (  \frac{\alpha}{0.1} \right )^{-2/3} 
 \left ( \frac {H/r} {10^{-3}} \right )^{-2/3} 
 \nn \\
 & \quad \times \nu_{500}^{2/3} M_{1.4}^{1/3} R_6^{1/3}. 
 \label{LTupper}
\end{align}
For example, even a mere $\theta=10^{\rm o}$ disc-spin misalignment would allow LT-induced tearing to take place
for $r/\Rs \lesssim 12$.

So far we have ignored the presence of the stellar magnetic field. Intuitively speaking, we would expect the field lines 
threading the disc to resist the tearing effect of the LT torque. This argument can be quantified if we compare the LT torque
with the local magnetic field torque. The latter parameter is given by $N_{\rm B} = r^3 B_\varphi B_z$ with the poloidal field given 
by Eq.~\eqref{polfield}. For the toroidal field we assume a functional form similar to the one of~\cite{wang95},
\be
B_\varphi (r)= B_z (r) f(\Oms/\OmK).
\ee
The magnetic field would be unable to prevent the tearing of the disc when $N_{\rm LT} \gtrsim N_{\rm B}$.
This condition is equivalent to
\be
\left ( \frac{r}{\Rs} \right )^{2}  \gtrsim \frac{15}{8}  \frac{ \alpha} { \sin\theta}  \frac{H}{r} 
\frac{ \Bs^2 \Rs}{ \cC \dot{M}  \Oms } |  f(\Oms/\OmK) |.
\label{ineq1}
\ee
Using normalised parameters (and approximating $f\sim 1$) we obtain the following numerical estimate
\begin{align}
\frac{r}{\Rs} &\gtrsim 0.7 (\sin\theta)^{-1/2} \left ( \frac{\alpha}{0.1} \right )^{1/2} \left ( \frac{H/r}{10^{-3}} \right )^{1/2} 
\nn \\
&\quad \times B_8 R_6 M_{1.4}^{-1/2}  \nu_{500}^{-1/2}  \dot{M}_{-10}^{-1/2}.
\label{LTlower}
\end{align}
This result leaves the door open for a possible LT-driven tearing of a magnetically-threaded disc.
For the previous example of a $\theta=10^{\rm o}$ misalignment (and for the rest of the parameters set to their
canonical values) we find $r \gtrsim 1.7 \Rs$.  Combined with the viscous upper limit~\eqref{LTupper},
our estimates suggest that the LT torque could play an important role in the dynamics of the inner part 
of accretion discs in LMXBs.  

To what extent LT precession could cause a fully non-linear fragmentation of the disc cannot be answered by the present
analysis. Numerical simulations of accreting black holes~\citep{nixon12,raj21} suggest that the formation of precessing 
`rings' takes place above an initial misalignment $\theta \approx 50^{\rm o}$; moreover, the emergence of this structure is 
accompanied by a markedly enhanced accretion rate due to loss of angular momentum between the orbiting rings. 
Order-of-magnitude variations in $\dot{M}$ during a burst, which may not be properly accounted for when averaging in the way 
described in Sec. \ref{sec:obs}, are likely to adjust the inferred parameters of the system. 

\section{Conclusions}
\label{sec:conclusions}

The main purpose of this paper is the comparison of the spin-up rates of a handful of AMXPs, with reliable timing data
during periods of burst activity and quiescence, against a collection of theoretically predicted accretion torque models.
Our results can be summarised as follows: (i) In all cases considered, none of the standard torque models endowed 
with magnetic field-disc coupling are able to explain the magnitude of the observed spin-up rates [in agreement with the 
findings of~\cite{ajh14}]; (ii) Thanks to their enhanced magnitude for $\xi<1$ (where $\xi$ is the phenomenological parameter 
that encapsulates much of the uncertain magnetospheric radius physics), the `new' torques devised in this paper 
(see Table~\ref{tab:accmodels}) predict spin-up rates comparable to the observed ones and at the same time are 
compatible with the systems' inferred dipole magnetic field strengths. Taking these results at face value we can conclude that, 
within the framework of standard accretion disc physics, the observed spin-up episodes in the examined AMXPs require 
$\xi \approx 0.1-0.5$; see Figs. \ref{fig:xte_xi} and \ref{fig:igr_xi}.

Moving beyond the standard accretion torque models, we have provided a quantitative analysis of the impact of some 
key additional physics effects. The inclusion of a quadrupole magnetic field component results in a flatter torque profile
as a function of the magnetic field strength but has only a moderate effect on the maximum torque. 
The inclusion of GR gravity leads to moderate corrections to the disc's orbital motion \citep{blanc98} and magnetospheric radius. 
As expected, the deviation from the Newtonian model diminishes (grows) with an increasing (decreasing) magnetic field 
as a result of the outwardly (inwardly) displaced magnetospheric radius. A perhaps more dramatic effect may take place,
driven by the action of the LT precession torque; if sufficiently inclined, the inner part of the disc might suffer a large scale 
fragmentation in spite of the cohesive counter-action of the viscous and magnetic forces, leading to huge variations in $\dot{M}$ 
over relatively short timescales \citep{nixon12,raj21}.

Not surprisingly, observational errors are part of life when it comes to modelling highly transient systems 
like AMXPs. A case in point is SAX J1808 with its multiply inferred dipole magnetic field strength during
periods of quiescence (see Fig.~\ref{fig:sax}). In a similar fashion, upper limit spin-up measurements (as in the case of 
XTE J1814 and IGR J17494) are rather poor probes of accretion torque physics and for that reason the aforementioned 
systems have been omitted from our analysis. The advent of new technologies such as NICER will undoubtedly improve the quality 
of future timing data, and may also be able to capture the spectroscopic evolution of emission lines in bright systems, which can be 
used as direct and independent probes for the inner radius of the accretion disc \citep{papp09,cack09}.

Taking the accretion torque modelling to the next level will probably require a shift from the analytical-phenomenological 
models discussed here to the full armoury of 3D numerical simulations [see, e.g., \cite{kulk13}]. The existing MHD codes, 
although still limited in terms of simulation time, have now reached a point where they can evolve an accretion flow without 
any symmetry imposed between the spin, disc and magnetic field axes \citep{rom20}. The numerical results
could serve as a test of the key ingredients of the phenomenological models such as the magnetospheric radius~\citep{kulk13} 
and the functional form of the generated azimuthal magnetic field [cf. Eq.~\eqref{Btor}; \citep{wang95,psaltis99}]. If robust enough, 
these results could be converted into analytical fit formulae and fed back into the phenomenological torque models.

\section*{Acknowledgements}
AGS gratefully acknowledges financial support from the Alexander von Humboldt Foundation. 


\section*{Data availability statement}
Observational data used in this paper are quoted from the cited works. 
Data generated from computations are reported in the body of the paper. 
Additional data can be made available upon reasonable request.



\appendix


\section{Thin-disc structure equations}
\label{sec:thindisc}

This short appendix summarises the textbook equations describing the structure of the standard Shakura-Shunyaev 
$\alpha$-viscosity thin disc model~\cite{accbook}. These equations are,
\be
\nu = \alpha c_s H, \quad c_s = H \Omega, \quad  \Sigma = \rho H,
\label{discparams}
\ee
where $H, \rho, \Sigma$ are, respectively, the disc's thickness,  density and surface density; 
$\Omega$ is the angular frequency, $c_s$ is the local `vertical' sound speed and $\nu$ is the shear 
viscosity coefficient. Viscosity is expressed in terms of the phenomenological $\alpha$ parameter. 

Another key equation of the model is the relation between surface density and accretion rate:
\be
\Sigma \approx \frac{\dot{M}}{3\pi\nu}.
\ee
For a Keplerian disc, this relation allows us to express the disc's density profile as,
\be
\rho(r) \approx  \frac {\dot{M}}{3\pi \alpha}   \left (\frac{H}{r} \right )^{-3} ( G\Ms r^3 )^{-1/2}.
\ee


\label{lastpage}

\end{document}